\documentclass{emulateapj}

\usepackage{amsmath}

\def\Ha{{\rm H}\alpha}
\def\HI{{{\rm HI}}}
\def\H2{{{\rm H}_2}}

\def\CO{{\rm CO}}
\def\UMW{U_{\rm MW}}
\def\MW{{\rm MW}}

\journalinfo{}
\submitted{submitted to APJ}

\shorttitle{The molecular hydrogen -- star formation relation}
\shortauthors{Feldmann, Gnedin \& Kravtsov}

\begin{document}

%=================================================
\title{The X-Factor in galaxies: II.  The molecular hydrogen -- star formation relation}
%=================================================

\author{Robert Feldmann\altaffilmark{1,2},
Nickolay Y. Gnedin\altaffilmark{1,2,3} and Andrey V. Kravtsov\altaffilmark{2,3,4}}  
\altaffiltext{1}{Particle Astrophysics Center, 
Fermi National Accelerator Laboratory, Batavia, IL 60510, USA; gnedin@fnal.gov}
\altaffiltext{2}{Kavli Institute for Cosmological Physics and Enrico
  Fermi Institute, The University of Chicago, Chicago, IL 60637 USA;
  andrey@oddjob.uchicago.edu} 
\altaffiltext{3}{Department of Astronomy \& Astrophysics, The
  University of Chicago, Chicago, IL 60637 USA} 
\altaffiltext{4}{Enrico Fermi Institute, The University of Chicago,
Chicago, IL 60637}

\begin{abstract}
There is ample observational evidence that the star formation rate (SFR) surface density, $\Sigma_{\rm SFR}$, is closely correlated with the surface density of molecular hydrogen, $\Sigma_\H2$. This empirical relation holds both for galaxy-wide averages and for individual $\gtrsim{}$kpc sized patches of the interstellar medium (ISM), but appears to degrade substantially at a sub-kpc scale.
Identifying the physical mechanisms that determine the scale-dependent properties of the observed $\Sigma_\H2-\Sigma_{\rm SFR}$ relation remains a  challenge from a theoretical perspective.
To address this question, we analyze the slope and scatter of the $\Sigma_\H2-\Sigma_{\rm SFR}$ relation using a set of cosmological, galaxy formation simulations with a peak resolution of $\sim{}100$ pc. These simulations include a chemical network for molecular hydrogen, a model for the $\CO$ emission, and a simple, stochastic prescription for star formation that operates on $\sim{}100$ pc scales.
Specifically, star formation is modeled as a Poisson process in which the average SFR is directly proportional to the present mass of $\H2$.
The predictions of our numerical model are in good agreement with the observed Kennicutt-Schmidt and $\Sigma_\H2-\Sigma_{\rm SFR}$ relations. We show that observations based on $\CO$ emission are ill suited to reliably measure the slope of the latter relation at low ($\lesssim{}20$ $M_\odot$ pc$^{-2}$) $\H2$ surface densities on sub-kpc scales. Our models also predict that the inferred $\Sigma_\H2-\Sigma_{\rm SFR}$ relation steepens at high $\H2$ surface densities as a result of the surface density dependence of the $\CO/\H2$ conversion factor. Finally, we show that on sub-kpc scales most of the scatter in the relation is a consequence of discreteness effects in the star formation process. In contrast, variations of the $\CO/\H2$ conversion factor are responsible for most of the scatter measured on super-kpc scales.
\end{abstract}


%----------------------
\section{Introduction}
\label{sec:intro}
%----------------------

The formation of stars is far from being a well understood and solved problem. The complex interplay between star formation, the formation of molecular clouds, and the physical processes operating within the interstellar medium (ISM), such as turbulence, magnetic fields, self-gravity, or feedback from the stellar population, poses many formidable challenges to the proper interpretation and modeling of the star formation process (e.g., \citealt{2007ARA&A..45..565M}).
Fortunately, observations have revealed a number of empirical relations which provide some guidance for the development of a theoretical model. This includes the well-studied Kennicutt-Schmidt relation, a correlation between the surface density of the neutral ISM and $\Sigma_{\rm SFR}$ \citep{1959ApJ...129..243S, 1989ApJ...344..685K, 1998ApJ...498..541K}. The realization that star formation is more closely correlated with molecular gas than neutral gas (\citealt{1980ApJ...235..821T, 2002ApJ...569..157W}, but cf. \citealt{1989ApJ...344..685K}) led to considerable interest in the study of the $\Sigma_\H2-\Sigma_{\rm SFR}$ relation (e.g., 
\citealt{2004ApJ...602..723H, 2004ApJ...606..271G, 2005MNRAS.359.1165G, 2007ApJ...669..289K, 2007ApJ...671..333K, 2008AJ....136.2846B, 2008ApJ...680.1083R, 2009ApJ...697...55G, 2009ApJ...699..850K, 2010ApJ...714..287G, 2011ApJ...728...88G, 2011ApJ...730L..13B, 2011ApJ...732..115F, 2011AJ....142...37S, 2011ApJ...730...72R, 2012ApJ...745..183R}).

The properties of the $\Sigma_\H2-\Sigma_{\rm SFR}$ relation are tied to the small scale (scales of molecular clouds and below) connection between star formation and the supply in form of molecular gas. Therefore, measuring the slope and scatter\footnote{This is a somewhat imprecise, but common, terminology. Here (and in the rest of the paper) ``slope'' and ``scatter'' of the $\Sigma_\H2-\Sigma_{\rm SFR}$ relation refer, more precisely, to the slope and to the standard deviation of the residuals (estimated from a linear regression) of the $\log_{10}\Sigma_\H2-\log_{10}\Sigma_{\rm SFR}$ relation.} of the $\Sigma_\H2-\Sigma_{\rm SFR}$ relation allows to probe potentially relevant physical mechanisms that are involved in the star formation process. This requires, of course, that biases in the observational tracers used to derive $\Sigma_\H2$ and $\Sigma_{\rm SFR}$ are properly accounted for. The main goal of this paper is therefore to study and isolate the importance of both various physical mechanisms and observational biases for the slope and scatter of the $\Sigma_\H2-\Sigma_{\rm SFR}$ relation.

Our theoretical predictions are based on high resolution, cosmological galaxy formation simulations equipped with subgrid models to estimate $\H2$ abundances and SFRs. We study observational biases by post-processing our simulations with a model for the $J=1\rightarrow{}0$ line emission of $\CO$ (\citealt{2012ApJ...747..124F}; paper I from now on). $\CO$ line emission is the most commonly used tracer to measure $\H2$ abundances (e.g., \citealt{1970ApJ...161L..43W, 1987ASSL..134...21S, 1991AJ....102.1956B, 1995ApJS...98..219Y, 2001ApJ...561..218R, 2003ApJS..145..259H, 2005ARA&A..43..677S, 2006ApJ...640..228T, 2006ApJ...650..604R, 2007PASJ...59..117K, 2010Natur.463..781T, 2010MNRAS.407.2091G, 2010ApJ...713..686D, 2010ApJ...714L.118D, 2011MNRAS.412.1913I}). Detecting $\H2$ in emission is difficult due to the high energy gap between the ground state and the lowest excited levels of $\H2$ and the lack of a permanent electric dipole moment of the $\H2$ molecule. Our $\CO$ emission model allows us to determine the $\CO/\H2$ conversion factor (also called the X-factor or $X_\CO$) which is formally defined as 
\begin{equation}
X_\CO = \frac{N_\H2}{W_\CO},
\end{equation}
i.e., as the ratio between $\H2$ column density, $N_\H2$, and the $\CO$ velocity integrated intensity, $W_\CO$, of the $J=1\rightarrow{}0$ rotational transition. As shown in paper I, the X-factor depends on the conditions of the ISM, in particular on the dust-to-gas ratio and the gas surface density. Hence, a crucial question that we address in this paper is whether the use of a constant value for $X_\CO$, as done in many observational studies, introduces significant biases in the estimates of the slope and scatter of the $\Sigma_\H2-\Sigma_{\rm SFR}$ relation. Similarly, we use stellar population synthesis modeling to compute the FUV and $\Ha$ fluxes of our model galaxies in order to assess whether the use of these star formation tracers leads to observational biases.

From a theoretical perspective, the slope of the $\Sigma_\H2-\Sigma_{\rm SFR}$ relation provides us with a means to distinguish between (and potentially rule out)  different models of star formation. In general, the slopes of the $\Sigma_\H2-\Sigma_{\rm SFR}$ relation and that of the underlying relation between the spatial densities of SFRs and $\H2$ \citep{1978A&A....68....1G, 1980ApJ...235..821T}, the $\rho_\H2-\dot{\rho}_*$ relation, could be different due to variations in the gas scale height \citep{1971ApL.....8..111T} or changes in the density distribution \citep{2011ApJ...732..115F}. However, if the $\rho_\H2-\dot{\rho}_*$ relation is linear, so is the $\Sigma_\H2-\Sigma_{\rm SFR}$ relation \citep{2008MNRAS.383.1210S, 2011ApJ...732..115F}. 

We note that a linear relation has been interpreted as evidence that star formation occurs in dense clumps within molecular clouds and that the number of such clumps scales with the total amount of molecular gas \citep{2008AJ....136.2846B}. An alternative possibility that has been suggested is that the timescale for star formation is controlled by small scale processes, e.g., stellar feedback, ambipolar diffusion, etc. which do not vary much from one location to the next, and hence ensure a steady (when averaged over sufficiently large spatial and temporal scales) conversion of $\H2$ into stars \citep{2002ApJ...569..157W}. In contrast, a slope of $\sim{}1.3$ can be explained by models that are based on a constant star formation efficiency per free-fall time \citep{2005ApJ...630..250K, 2009ApJ...699..850K}. Star formation driven by cloud-cloud collisions could lead to  even steeper slopes \citep{2011ApJ...730...11T}.

While the slope of the $\Sigma_\H2-\Sigma_{\rm SFR}$ relation has been the subject of a number of observational studies, there is, unfortunately, no clear consensus yet on its exact value. While many studies find slopes that are close to unity ($\sim{}0.8-1.3$, e.g., \citealt{2002ApJ...569..157W, 2005PASJ...57..733K, 2008AJ....136.2846B, 2009ApJ...704..842B, 2011ApJ...730L..13B, 2011AJ....142...37S, 2011ApJ...730...72R, 2012ApJ...745..183R}), a number of works prefer a steeper slope, e.g., $\sim{}1.4$ \citep{2004ApJ...602..723H, 2007ApJ...671..333K}, $\sim{}1.6$ \citep{2007ApJS..173..572T}, $\sim{}1.2-1.9$ \citep{2011ApJ...735...63L}. Various factors may contribute to these differences, including choices in the fitting methodology, how star formation maps are corrected for dust extinction and diffuse emission, the resolution of the observations, or the range in surface densities over which the fit is done. In particular, assumptions about the amount of diffuse emission present in the star formation maps and its subsequent treatment in the data analysis have a strong impact on the derived slope \citep{2007ApJ...671..333K, 2011ApJ...735...63L, 2011ApJ...730...72R}. However, if the fitting is restricted to regions of sufficiently high $\Sigma_\H2$, then the impact of any diffuse emission is minimized and the $\Sigma_\H2-\Sigma_{\rm SFR}$ relation is found to be close to linear \citep{2012ApJ...745..183R}.

Besides the slope, the normalization of the $\Sigma_\H2-\Sigma_{\rm SFR}$ relation is an important parameter that needs to be matched by the theoretical models. Fortunately, studies that find an approximatively linear slope of the $\Sigma_\H2-\Sigma_{\rm SFR}$ relation also find little variation of the normalization ($\Sigma_\H2/\Sigma_{\rm SFR}$ $\sim{}$few Gyr) with galaxy mass or other global galaxy properties (e.g., \citealt{2001A&A...378...51B, 2002ApJ...569..157W, 2008AJ....136.2846B, 2010MNRAS.407.2091G, 2011ApJ...730L..13B, 2011ApJ...741...12B}). A notable exception is the claim that the normalization depends on the specific star formation rate \citep{2011MNRAS.415...61S}, although there is a potential worry that this result is, to some extent at least, an artifact of the large scatter in the $\Sigma_\H2-\Sigma_{\rm SFR}$ relation and the use of correlated quantities \citep{2012ApJ...745..183R}.
Driven by these observational findings we adopt in this paper a (stochastic) star formation model that is based on a linear $\rho_\H2-\dot{\rho}_*$ relation with a constant gas depletion time of a few Gyr.

\begin{table*}
\begin{center}
\caption{Details of the numerical simulations}
\begin{tabular}{lcccccr}
\tableline \tableline
label & box size & $\Delta{}x$ & $m_{\rm DM}$ & [$\Omega_{\rm m}, \Omega_\Lambda, \Omega_{\rm b}, h, \sigma_8$] & fixed ISM & comments \\ 
\tableline \noalign{\smallskip}
MW-fid   & 6 com. Mpc h$^{-1}$ & 65 pc & $9\times{}10^5$ $M_\odot$ h$^{-1}$ & [0.3, 0.7, 0.043, 0.7, 0.9] & yes & fiducial SF model\\
MW-dt6   & 6 com. Mpc h$^{-1}$ & 65 pc & $9\times{}10^5$ $M_\odot$ h$^{-1}$  & [0.3, 0.7, 0.043, 0.7, 0.9] & yes & $\Delta{}t_{\rm SF}=10^6$ yr \\
MW-dt5   & 6 com. Mpc h$^{-1}$ & 65 pc & $9\times{}10^5$ $M_\odot$ h$^{-1}$ & [0.3, 0.7, 0.043, 0.7, 0.9] & yes & $\Delta{}t_{\rm SF}=10^5$ yr \\
MW-sl2   & 6 com. Mpc h$^{-1}$ & 65 pc & $9\times{}10^5$ $M_\odot$ h$^{-1}$   & [0.3, 0.7, 0.043, 0.7, 0.9] & yes & slope-2 SF model\\
HZ-csm    & 25.6 com. Mpc h$^{-1}$ & 97 pc & $1\times{}10^6$ $M_\odot$  h$^{-1}$ & [0.28, 0.72, 0.046, 0.7, 0.82] & no & down to $z=1.8$ \\
HZ-fid     & 25.6 com. Mpc h$^{-1}$ & 97 pc & $1\times{}10^6$ $M_\odot$  h$^{-1}$ & [0.28, 0.72, 0.046, 0.7, 0.82] & yes & fiducial SF model\\
\tableline \tableline
\end{tabular}
\label{tab:sim}
\tablecomments{The first five columns provide the name, box size, numerical resolution, and the adopted cosmological parameters of each zoom-in simulation. The penultimate column states whether the simulation is run with gas-to-dust ratios and interstellar radiation fields fixed to the corresponding values in the solar neighborhood. The parameter $\Delta{}t_{\rm SF}$ in the last column refers to the average time between individual star formation events, see \S{}\ref{sect:SFmodel}. The fiducial SF model uses $\Delta{}t_{\rm SF}=10^7$ yr.}
\end{center}
\end{table*}

The scatter in the $\Sigma_\H2-\Sigma_{\rm SFR}$ relation has received significantly less attention compared to the slope or the normalization. The reason is probably that the scatter depends even more on the specific details of the respective observational survey and the data analysis procedures. The finding that the scatter increases with increasing spatial resolution of a survey is sometimes used to argue that star formation scaling relations ``break down'' on small scales \citep{2010ApJ...721..383M, 2010ApJ...722L.127O, 2010ApJ...722.1699S}. However, we will argue in this paper that the proper interpretation is that on small scales the intrinsically stochastic nature of the scaling relations becomes simply more evident. 

A first step to understand and quantify the different contributors to the scatter was made by \cite{2011ApJ...732..115F}. Here, we extend our previous analysis in several ways. We consider the scatter that results from spatial X-factor fluctuations. We also analyze the scatter related to star formation tracers ($\Ha$ and $FUV$). Finally, we highlight the importance of stochasticity in the star formation process and propose a simple model of star formation that allows a classification of the various mechanisms that contribute to the scatter in the $\Sigma_\H2-\Sigma_{\rm SFR}$ relation.

The outline of the paper is as follows. In \S\ref{sect:methods} we present the detail of our numerical approach, including the set-up of the simulations (\S\ref{sect:Sims}), the star formation model (\S\ref{sect:SFmodel}), details of sub-grid physics (\S\ref{sect:SubGridModeling}), and the modeling of the $\H2$ and star formation tracers (\S\ref{sect:PostProcessing}). We then show in \S\ref{sect:SFrelations} that our simulations are able to reproduce the observed Kennicutt-Schmidt and $\Sigma_\H2-\Sigma_{\rm SFR}$ relations. Next, in \S\ref{sect:SuperLinear}, we discuss observational claims of a non-linear  $\rho_\H2-\dot{\rho}_*$ relation based on slope measurements of the $\Sigma_\H2-\Sigma_{\rm SFR}$ relation at low $\Sigma_\H2$. Then, in \S\ref{sect:highZ} we analyze the importance of the X-factor for high redshift galaxies with high gas surface densities. We address the origin of the scatter in the $\Sigma_\H2-\Sigma_{\rm SFR}$ relation in \S\ref{sect:Scatter}. In \S\ref{sect:DiscussionScatter} we discuss our star formation model in the context of various observational constraints. Finally, in \S\ref{sect:Conclusions}, we summarize our results and conclusions.

%----------------------
\section{Methodology}
\label{sect:methods}
%----------------------

\subsection{Simulations}
\label{sect:Sims}

All simulations have been run with the Eulerian hydrodynamics + N-body code ART \citep{1997ApJS..111...73K, 2002ApJ...571..563K} that uses an adaptive mesh refinement (AMR) technique to increase the resolution selectively in specified regions of interest. We also use the standard method of embedding these regions in layers of subsequently lower dark matter resolution to further reduce the computational cost, but still capture the impact of large scale tidal fields correctly \citep{1991ApJ...368..325K, 2001ApJS..137....1B}. 

Simulation MW-fid focusses its computational resources on a Lagrangian region that encloses five virial radii of a MW-sized halo at $z=0$ (total mass $\sim{}10^{12}$ $M_\odot$) in a 6 Mpc h$^{-1}$ box. The simulation is started from cosmological initial conditions with the parameters given in  Table~\ref{tab:sim}. It is run fully self-consistently down to redshift $z=4$. At this point ``fixed ISM conditions'' are imposed on the simulation as described in \S\ref{sect:SubGridModeling} and it is continued for additional 600 Myr before it is analyzed.
By then the high-resolution Lagrangian region harbors a large disk galaxy with a virial mass of $\sim{}4.2\times{}10^{11}$ $M_\odot$ and several less massive galaxies.  More details on the setup of the MW-fid simulation can be found elsewhere (\citealt{2011ApJ...728...88G, 2011ApJ...732..115F}; paper I).

The simulations MW-dt5, MW-dt6, and MW-sl2 are spawned from the $z=4$ snapshot of MW-fid and continued for additional 600 Myr with fixed ISM conditions. The first two simulations are run with reduces values of $\Delta{}t_{\rm SF}$, the average time scale between individual star formation events, see \S{}\ref{sect:SFmodel}, but are otherwise identical to MW-fid. The run MW-sl2 uses a modified star formation model, see \S\ref{sect:SuperLinear} for details. 

Simulation HZ-csm refines on seven regions within a 25 Mpc h$^{-1}$ box. By $z=0$, these regions have collapsed to halos in the $10^{11}-10^{13}$ $M_\odot$ mass range. The simulation is started at $z=100$ and is continued down to $z=1.8$ fully self-consistently. The setup of the HZ-csm simulation is discussed in detail in \cite{2012ApJ...748...54Z}. Simulation HZ-fid is spawned from the $z=2$ snapshot of HZ-csm. It is continued with fixed ISM conditions for additional 200 Myr (down to $z=1.8$)  to allow the gas to react to the changes in the ISM properties and to reach its new equilibrium state. 

The main properties of our simulations are summarized in Table~\ref{tab:sim}.

\subsection{Star formation model}
\label{sect:SFmodel}

Since individual resolution elements in our simulations correspond to (at best) GMC scales, it is clear that we are not able to follow in any realistic detail the formation and evolution of individual bound star clusters, let along individual stars. Hence, our approach is to marginalize over the complexities involved in star formation by describing star formation on a statistical level. Such an approach will have its limitations, of course, and we are not aiming at reproducing observations perfectly, but rather try to isolate and explain some of the trends in the observational data. 

There are several lines of evidence suggesting that a statistical description of star formation is a plausible ansatz, at least on the level of star clusters and in relatively quiescently star-forming galaxies. One is the observation that the scaling between $\Sigma_{\rm SFR}$ and $\Sigma_\H2$ is roughly linear and holds over a wide range of surface densities and galaxy metallicities \citep{2008AJ....136.2846B, 2010MNRAS.407.2091G}, although it may break down in strongly out-of-equilibrium situations such a major galaxy mergers \citep{2010ApJ...720L.149T, 2011ApJ...730....4B}. Another one is provided by the similarity of the shape of the mass function of young star clusters among non-starbursting galaxies \citep{2010ARA&A..48..431P}, which indicates that the conversion of interstellar gas into star clusters proceeds on average in the same way in different galaxies with different global properties. Also, the existence of a correlation between the total number of clusters and the luminosity of the most luminous cluster in a galaxy can be understood in a purely statistical formation scenario of star clusters \citep{2002AJ....124.1393L}.

Our model assumes that  $\H2$ mass is a good tracer of the SFR \citep{2006ApJ...645.1024P, 2008ApJ...680.1083R, 2009ApJ...697...55G, 2010ApJ...717.1037P,2011ApJ...728...88G, 2011ApJ...732..115F, 2012ApJ...749...36K} and can be summarized by the following three statements: (1) star formation is a stochastic process and the number of individual star formation events\footnote{The proper interpretation of an individual star formation event is as the \emph{simultaneous} formation of a number of embedded star clusters in a given region in space ($\sim{}60-100$ pc in our simulations).} in a given time interval is described by a homogeneous Poisson process, (2) the \emph{average} stellar mass formed per unit time is proportional to the present $\H2$ mass, and (3) the factor of proportionality, $\tau_{\rm dep}^{-1}$, is a constant. Here, \emph{average} refers to the ensemble average over independent patches of the ISM of the same size and with the same $\H2$ masses, which may, however, differ in other properties. In other words, we marginalize over most of the GMC parameters which may be potentially relevant for star formation and keep only the explicit dependence of the SFR on $\H2$ mass.

We give a detailed description of the model in the appendix. Here, we only summarize the properties that we will need later in the paper. 

The discreteness of star formation leads to scatter in the $\Sigma_\H2-\Sigma_{\rm SFR}$ relation because the SFR that is realized in a given region and over a given period differs from the ensemble average SFR tied to the present $\H2$ mass. The scatter increases (up to a point) with increasing average time between individual star formation events, $\Delta{}t_{\rm SF}$, but decreases with increasing life-time of a given star formation tracer $\Delta{}t_*$. In fact, the scatter depends on these quantities only in form of the dimensionless ratio $\Delta{}t_*/\Delta{}t_{\rm SF}$, the average number of individual star formation events during $\Delta{}t_*$. 

Our model contains three parameters: the average $\H2$ depletion time scale, $\tau_{\rm dep}$, the spatial scale on which our model operates, $l_{\rm SF}$, and $\Delta{}t_{\rm SF}$. We fix $\tau_{\rm dep}$ to $2.9$ Gyr in order to fit the observed normalization of the $\Sigma_{\rm SFR}$ - $\Sigma_\H2$ relation, see \S\ref{sect:SFrelations}. We stress that $\tau_{\rm dep}$ is not the $\H2$ gas depletion time of individual star forming molecular clouds, but an average depletion time that includes the non star forming molecular gas. Our model is assumed to operate on scales of $l_{\rm SF}=60-100$ pc scales, which is the peak resolution of our numerical simulations, see Table~\ref{tab:sim}. Our default value for $\Delta{}t_{\rm SF}$ is 10 Myr. This choice is based on the following considerations. 

First, the crossing time (or the free-fall time) of a molecular region should be a lower limit on the time period between two bursts of star formation in that region. This time $t_{\rm cross}\sim{}L/\sigma$ can be estimated from the Larson relation between cloud size $L$ and cloud velocity dispersion $\sigma$ \citep{1981MNRAS.194..809L, 1987ApJ...319..730S}. It scales as $\sim{}1\,{\rm Myr}\,(L/{\rm pc})^{0.5}$ and, hence, $t_{\rm cross}\sim{}8-10$ Myr for regions of size $l_{\rm SF}=60-100$ pc. Since the largest molecular cloud complexes in the Milky Way have radii of $\sim{}100$ pc \citep{1986ApJ...305..892D}, this line of reasoning cannot be used to constrain $\Delta{}t_{\rm SF}$ on larger spatial scales. Despite having a similar numerical value on $\sim{}100$ pc scales, $\Delta{}t_{\rm SF}$ and $t_{\rm cross}$ are quite different physical quantities. In fact, $t_{\rm cross}$, a measure of the \emph{duration} of an individual star formation event, increases with scale, while $\Delta{}t_{\rm SF}$, the period between successive star formation events, decreases with scale due to the smoothing effect of spatial averaging on Poisson noise.

Secondly, we have checked that with the choice $\Delta{}t_{\rm SF}=10$ Myr individual star formation events in the simulations cover a broad range of masses up to $10^6$ $M_\odot$. The embedded star clusters that form during a star formation event of total mass $10^6$ $M_\odot$ will have masses up to, but not exceeding, $\sim{}10^5$ $M_\odot$ for a reasonable choice of the slope of the embedded cluster mass function ($\sim{}2.3-2.4$). This compares well with the fact that, at least in the Milky Way, young massive clusters exceeding $10^5$ $M_\odot$ are rare \citep{2010ARA&A..48..431P}. 

Finally, $\Delta{}t_{\rm SF}\sim{}10$ Myr is consistent with the observed relation between SFR and maximum embedded star cluster mass \citep{2004MNRAS.350.1503W}. We note that these considerations do not rule out a moderately larger value, e.g, $\Delta{}t_{\rm SF}\sim{}20$ Myr.

\subsection{Further sub-grid modeling}
\label{sect:SubGridModeling}

All simulations include a photo-chemical network, metal enrichment from supernova (type Ia and type II), but no thermal energy injection, optically thin radiative cooling by hydrogen (including $\H2$), helium, and metal lines, and 3D radiative transfer of UV radiation. The details of the implementation can be found in \cite{2009ApJ...697...55G} and \cite{2011ApJ...728...88G}. Here, we give a brief recount.

The photo-chemical network in ART follows the formation and destruction of the five major atomic and ionic species of hydrogen and helium. The formation of $\H2$ on dust grains and the destruction of $\H2$ via photo-dissociation in the Lyman-Werner bands are taken into account. The transfer of ionizing and non-ionizing UV radiation from stellar sources is computed in the OTVET approximation \citep{2001NewA....6..437G}. Radiative stellar feedback is important for the heating and cooling balance of the gas and for the abundance of $\H2$ and CO. Unlike the re-ionized intergalactic medium, the dense interstellar medium within galaxies may well be opaque to ionizing photons of all but the nearest stars and radiative transfer effects cannot be neglected.

\begin{figure*}
\begin{tabular}{cc}
\includegraphics[width=80mm]{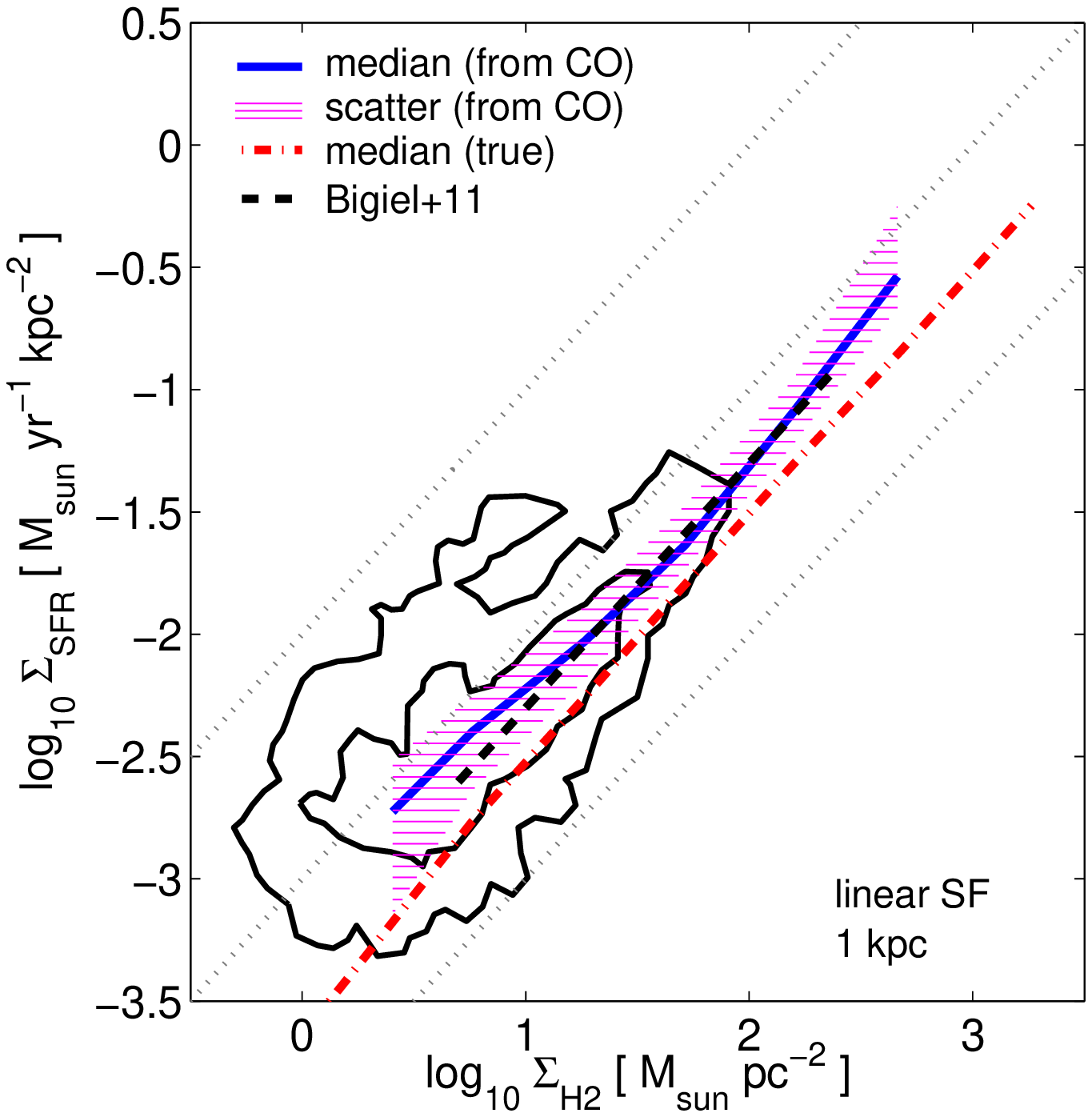} &
\includegraphics[width=80mm]{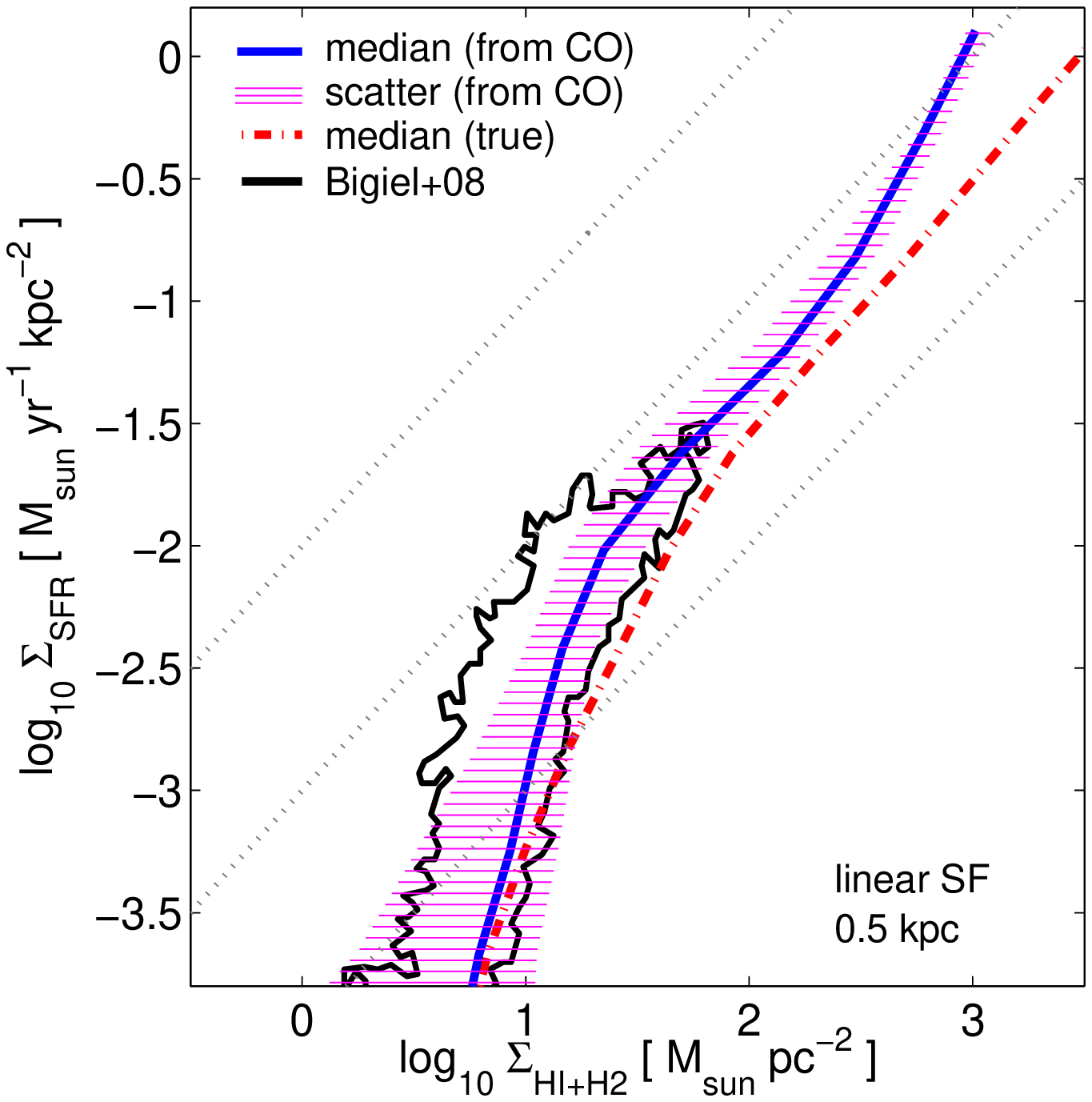} \\
\end{tabular}
\caption{Relation between the molecular gas (neutral gas) surface density $\Sigma_\H2$ ($\Sigma_{\HI+\H2}$) and the surface density of the star formation rate $\Sigma_{\rm SFR}$ as predicted by our fiducial SF and CO emission model for Milky-Way like ISM properties ($D_\MW=1$, $\UMW=1$).
The left panel shows the $\Sigma_\H2$ vs $\Sigma_{\rm SFR}$ relation, the right panel the classical Kennicutt-Schmidt relation ($\Sigma_{\HI+\H2}$ vs $\Sigma_{\rm SFR}$). The spatial resolution is chosen to match as closely as possible the resolution of the observational studies \citep{2008AJ....136.2846B, 2011ApJ...730L..13B}. The gas surface densities include a factor 1.36 that accounts for the presence of Helium.
The red dot-dashed line shows the median $\Sigma_{\rm SFR}$ (time averaged over the past 20 Myr) for a given ``true'' $\H2$ or total gas surface density (as computed in the simulation). The blue solid line shows instead the median of $\Sigma_{\rm SFR}$ as a function of the ``inferred'' $\H2$ or total gas surface density,
i.e., a gas density in which $\Sigma_\H2$ is inferred from the $\CO$ intensity, as predicted by our model, using the galactic X-factor. The magenta hashed region indicates the typical scatter in $\Sigma_{\rm SFR}$ (25 and 75-th percentiles in the left panel, 16 and 84-th percentiles in the right panel, ). The black dashed line (best fit) and the contour lines (containing 90\% and 50\% of the data) in the left panel are observational measurement of the $\Sigma_\H2$ - $\Sigma_{\rm SFR}$ relation by \cite{2011ApJ...730L..13B} on kpc scales. The contour line in the right panel shows the distribution of sub-kpc sized patches in the sample of nearby galaxies by \cite{2008AJ....136.2846B}, see their Fig. 7. The figure demonstrates that our modeling of the star formation and the CO emission is consistent with the observed relations between $\Sigma_{\rm SFR}$ and the molecular and neutral gas surface density, respectively.}
\label{fig:KS}
\end{figure*}

Most of our simulations are run with ``fixed ISM conditions'' (see \citealt{2011ApJ...728...88G}). First of all this means that the dust-to-gas ratio $D_\MW$, which is normally assumed to scale linearly with the local gas metallicity, is kept fixed at a value that corresponds to\footnote{In this paper $Z_\odot$ refers to the metallicity of the solar neighborhood. Specifically, $Z_\odot=0.02$ or 12 + $\log_{10}$(O/H) = 8.92, which is somewhat larger than the metallicity of the Sun according to recent estimates \citep{2001ApJ...556L..63A, 2004A&A...417..751A, 2009ARA&A..47..481A}.} $Z_\odot$, and we will denote this as $D_\MW=1$. The gas-to-dust ratio is a crucial parameter that enters the formation rates and the dust shielding of molecular hydrogen and also our CO emission model. The metallicities of the self-consistently enriched gas are still used to compute the cooling rates that enter the hydrodynamical solver. This is done in order to avoid numerical artifacts such as sudden increases in the gas accretion rates resulting from excess, non-equilibrium cooling. 

Furthermore, in the fixed ISM runs, the normalization of the radiation field at 1000 {\AA} is fixed to $J_{\rm MW}$ = 10$^6$ photons cm$^{-2}$ s$^{-1}$ sr$^{-1}$ eV$^{-1}$, a value typical for the solar neighborhood in the Milky Way \citep{1978ApJS...36..595D, 1983A&A...128..212M}. We use the notation $\UMW=1$, where $\UMW$ is the intensity of the radiation field at 1000 {\AA} in units of $J_{\rm MW}$. We stress that only the normalization of the radiation field computed with the OTVET solver is fixed. The shape of the radiation spectrum is not modified.

\subsection{Postprocessing: CO, FUV and $\Ha$ emission}
\label{sect:PostProcessing}

The $J=1\rightarrow{}0$ $^{12}\mathrm{C}^{16}\mathrm{O}$ emission is computed as described in paper I. In short, a sub-grid model for the CO emission is constructed based on the results of a suite of small scale magneto-hydrodynamical ISM simulations \citep{2011MNRAS.412..337G}. This model contains two free parameters. 

One is the $\CO$ brightness temperature that is related to both the temperature of the $\CO$ emitting gas and the temperature of the cosmic microwave background (CMB). In \S\ref{sect:SFrelations}, \S\ref{sect:SuperLinear}, and \S\ref{sect:Scatter} we study the $\Sigma_\H2$ - $\Sigma_{\rm SFR}$ in the local Universe and we adopt a gas temperature of 10 K (a typical temperature of molecular clouds in the Milky Way). The corresponding brightness temperature is 6.65 K. In \S\ref{sect:highZ} we predict the $\CO$ emission for galaxies at $z\sim{}2$. We assume that the increase in the CMB temperature at those redshifts is compensated by an increase in the gas temperature from $10$ K to $14.5$ K, such that the brightness temperature remains approximatively constant. Such a moderate increase in gas temperature is consistent with the detailed modeling of the gas temperature in non-starbursting high redshift galaxies using photon-dissociation regions codes coupled with a semi-analytic galaxy evolution model \citep{2012arXiv1204.0795L}.

The other free parameter is the scaling of the $\CO$ line width (either a constant line width or a virial scaling). By default we show results for the virial line width scaling, which is probably the more realistic of the two (see Fig. 3 of paper I), but if relevant we will point out the changes that result from the assumption of a constant line width. The sub-grid model is applied to the highest refined resolution elements in the simulation. The contributions from these individual, $\sim{}60-100$ pc sized resolution elements are then combined in the optically thin limit to derive the CO emission from larger regions. 

The FUV and $\Ha$ emission of each stellar particle is computed with Starburst-99 \citep{2010ApJS..189..309L} assuming solar metallicity, high mass loss Geneva tracks. We checked that switching to, e.g., Padua tracks does not affect any of our results in a significant way. The $\Ha$ luminosities are directly taken from the Starburst-99 output, while the broad-band FUV luminosities are computed using the Galex FUV transmission curve \citep{2005ApJ...619L...7M} and the UV spectra provided by Starburst-99.

\begin{figure*}
\begin{tabular}{cc}
\includegraphics[width=80mm]{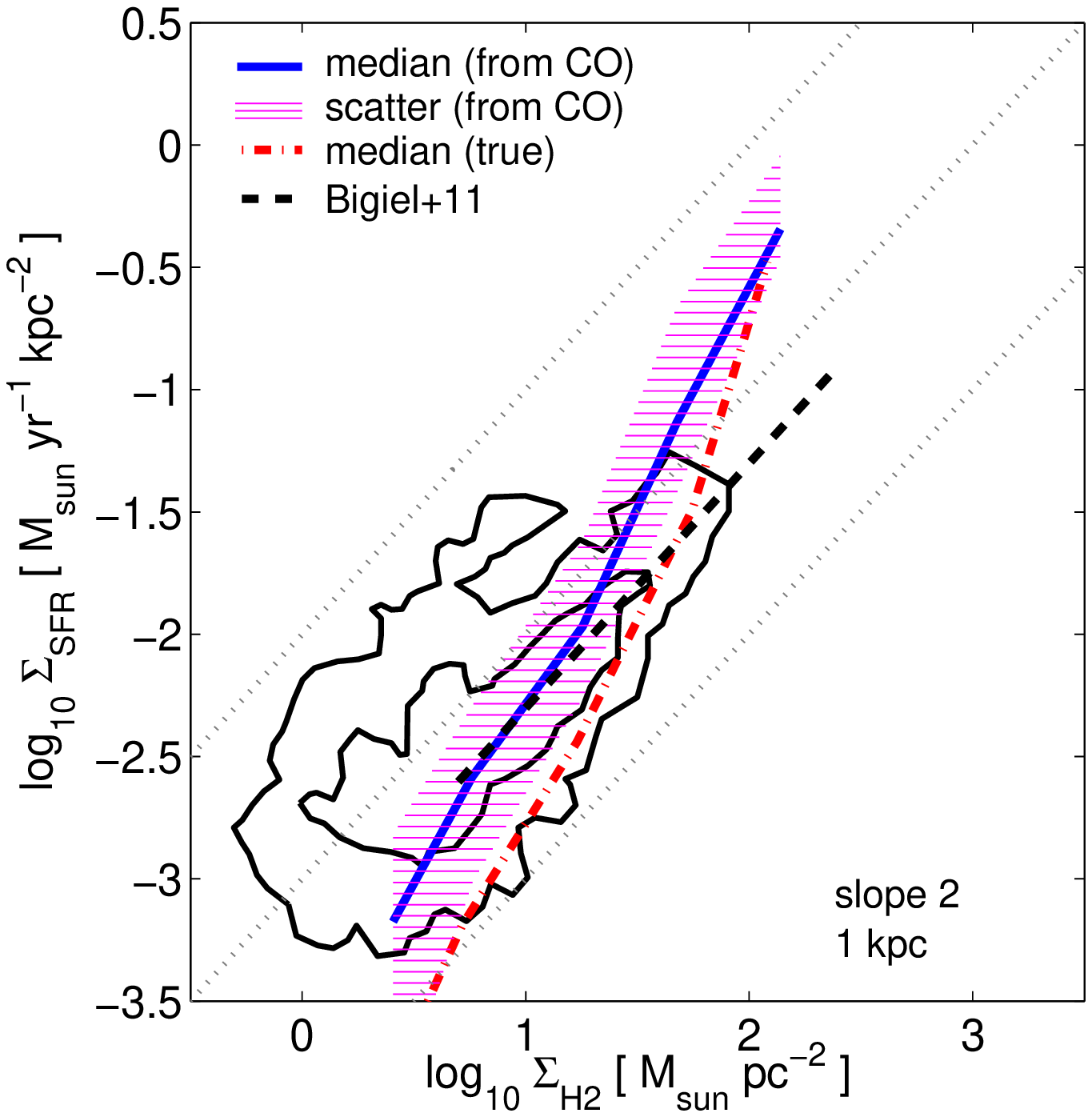} &
\includegraphics[width=80mm]{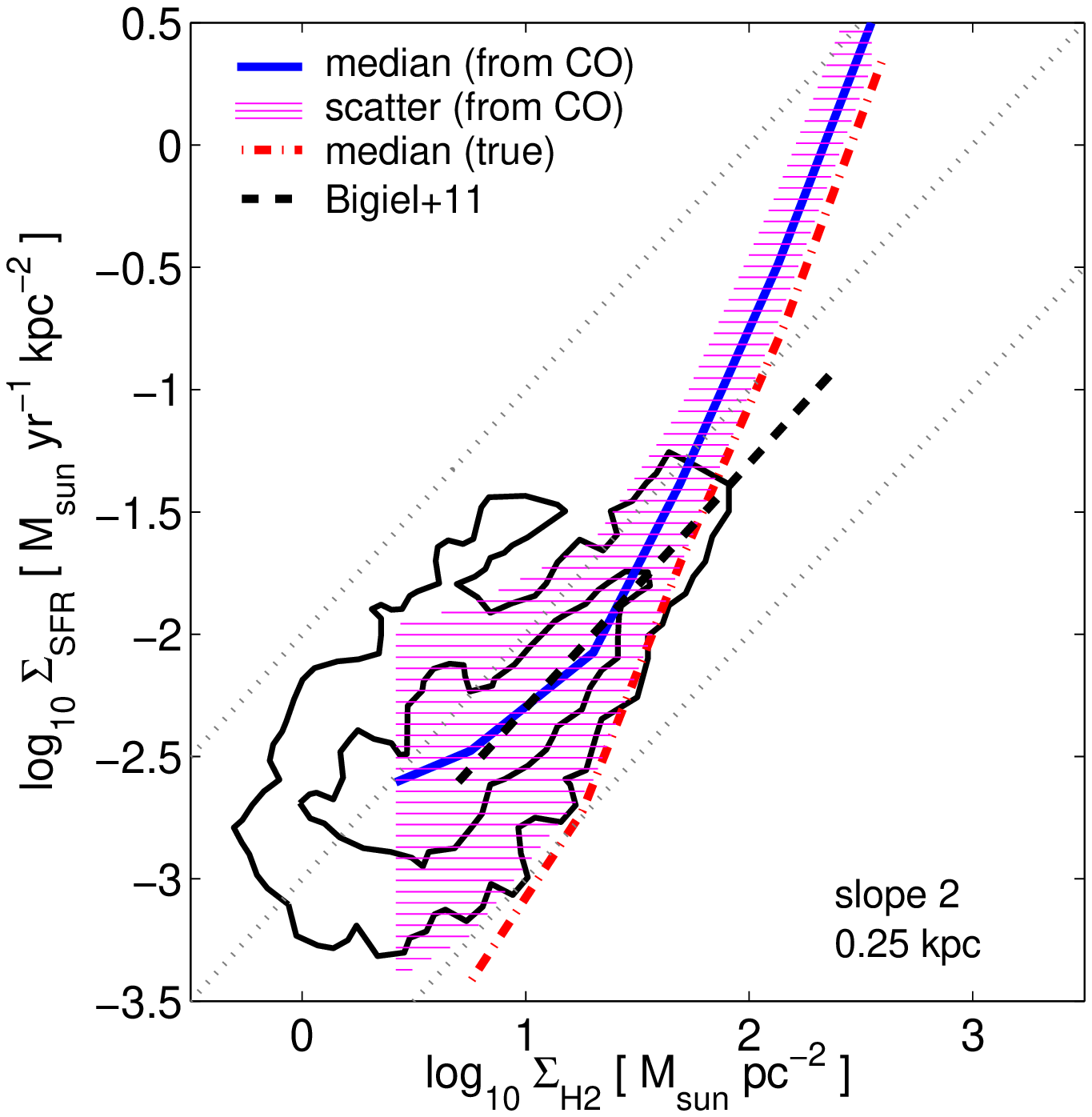} \\
\end{tabular}
\caption{Same as the left panel in Fig.\ref{fig:KS}, but for a SF model in which $\dot{\rho}_*\propto{}\rho_\H2^2$. The left panel uses surface densities measured on kpc scales, while the right panel shows the corresponding results for spatial averaging scales of 250 pc. This SF model leads to a steepening of the inferred $\Sigma_\H2-\Sigma_{\rm SFR}$ relation at sufficiently high $\H2$ surface densities ($\gtrsim{}20$ $M_\odot$ pc$^{-2}$). However, this steepening is suppressed for $\Sigma_\H2 < 20$ $M_\odot$ pc$^{-2}$, especially when measured at sub-kpc spatial resolution, as a result of the increase of $X_\CO$ with decreasing $\Sigma_\H2$ at low surface densities (paper I). It will therefore be challenging to extract an unbiased estimate for the slope of the relation between SFR and gas density from CO observations of MW-like galaxies at low surface densities.}
\label{fig:KS2}
\end{figure*}

%--------------------
\section{Results}
%--------------------

\subsection{Actual and inferred star formation relations}
\label{sect:SFrelations}

It has been shown \citep{2009ApJ...697...55G, 2009ApJ...699..850K, 2010ApJ...714..287G} that an $\H2$-based star formation prescription is able to reproduce the relation between the surface densities of neutral gas and SFRs (the Kennicutt-Schmidt relation) both for galaxies in the local Universe and for galaxies at high redshift. In fact, the connection between star formation and molecular hydrogen provides a simple physical interpretation for the drop in the SFR surface densities at low gas surface densities \citep{2008ApJ...680.1083R}. In this picture the drop is a manifestation of a transition between a neutral and a molecular hydrogen phase. Specifically, above a characteristic gas surface density the gas is shielded from the interstellar radiation field by a sufficiently large dust optical depth and the molecular phase prevails. The characteristic surface density depends on the dust-to-gas ratio, the strength of the interstellar radiation field, and the density structure of the ISM, see, e.g, \cite{2011ApJ...728...88G}.

However, this interpretation of the Kennicutt-Schmidt relation neglects a potentially relevant detail. While theoretical models can predict $\H2$ surface densities directly, observations often rely on $\CO$ observations to infer $\H2$ surface densities. A crucial test of our understanding of the Kennicutt-Schmidt relation is therefore whether this agreement still holds even if we take the effects of the CO/$\H2$ conversion factor into account. In other words, we need to compare theory and observations on an equal footing, i.e., using $\H2$ surface densities that are derived from $\CO$ emission in both cases.

Our star formation model contains the gas depletion time $\tau_{\rm dep}$ as a free parameter. This parameter is the conversion factor between $\H2$ mass and the ensemble average SFR, see \S\ref{sect:SFmodel}, but, since we assume $\tau_{\rm dep}=$const, it also corresponds to the normalization of the $\Sigma_\H2-\Sigma_{\rm SFR}$ relation. We find that a depletion time $\tau_{\rm dep}$ of 2.9 Gyr leads to an inferred ($\CO$ based) depletion time of $\sim{}2.35$ Gyr (including Helium, \citealt{2011ApJ...730L..13B}) and, thus, to excellent agreement between predictions and observations, see Fig.~\ref{fig:KS}a. We note that the proper choice of $\tau_{\rm dep}$ is largely degenerate with the value of the X-factor. Specifically, our $\CO$ emission model prefers a median $X_\CO$ value for a Milky Way like ISM that that is $\sim{}65\%$ larger than the value $X_{\CO,{\rm MW}}=2\times{}10^{20}$ cm$^{-2}$ K$^{-1}$ km$^{-1}$ s used by \cite{2011ApJ...730L..13B}, explaining the need for a somewhat larger $\H2$ depletion time.

Fig.~\ref{fig:KS}a shows the inferred ($\CO$ based) $\Sigma_\H2-\Sigma_{\rm SFR}$ relation based on our simulation MW-fid.
It is an approximatively linear relation over two orders of magnitude in $\H2$ surface density. This is not an entirely obvious result since, at least on sufficiently small scales, the $\CO$/$\H2$ conversion factor is strongly dependent on surface density (paper I). However, it turns out that the spatial averaging over a large set of regions with different $X_\CO$ values erases most of this surface density dependence on kpc and larger scales. Hence, the actual (linear) slope of the $\Sigma_\H2-\Sigma_{\rm SFR}$ relation is recovered (but see \S\ref{sect:highZ}).

We show our predictions for the Kennicutt-Schmidt relation in Fig.~\ref{fig:KS}b, finding good agreement between our theoretical predictions and observations of the inferred ($\CO$ based) Kennicutt-Schmidt relation. We thus conclude that a star formation model that depends linearly on the abundance of $\H2$ is consistent with both the observed Kennicutt-Schmidt relation and the observed $\Sigma_\H2-\Sigma_{\rm SFR}$ relation of normal star-forming galaxies in the local Universe.

\subsection{The slope of the $\Sigma_\H2-\Sigma_{\rm SFR}$ relation}
\label{sect:SuperLinear}

The slope of the $\Sigma_\H2-\Sigma_{\rm SFR}$ relation has been a subject of significant debate over the last years. While a number of studies find a slope of about unity, other observational works favor steeper slopes. It has been argued that these discrepancies arise from observational obstacles such as diffuse emission in the infrared, or uncertainties in the dust absorption that make it challenging to obtain reliable SFR estimates, especially in regions of low star formation activity and gas surface density \citep{2011ApJ...735...63L, 2011ApJ...730...72R, 2012arXiv1202.2873L}.

A varying $\CO/\H2$ conversion factor is another of those potential obstacles, but has been largely neglected since there is no general observational handle on this quantity in different galaxies. The aim of this section is to test the implications for the observed $\Sigma_\H2-\Sigma_{\rm SFR}$ relation if the $\dot{\rho_*}-\rho_\H2$ relation is highly non-linear and the X-factor is taken into account. In the following we assume that any observational uncertainties related to the estimates of SFR  (e.g., diffuse infrared emission, dust absorption, etc.) are under control and corrected for. We focus instead on the role of the $\CO/\H2$ conversion factor.

For this test we use simulation MW-sl2, which uses the same general approach to star formation as presented in \S\ref{sect:SFmodel}, but with a density dependent
gas depletion time. Specifically, we assume that $\tau_{\rm dep}\propto{}1/\rho_\H2$ which implies that (on average) $\dot{\rho_*}\propto{}\rho_\H2^2$. The factor of proportionality is chosen such that the gas depletion time, as estimated from the inferred ($\CO$ based) $\Sigma_\H2-\Sigma_{\rm SFR}$ relation measured on kpc scales, is $\sim{}2.3$ Gyr at $\Sigma_\H2=10$ $M_\odot$ pc$^{-2}$.

The results of this test are shown in Fig.~\ref{fig:KS2}. While the actual $\Sigma_\H2-\Sigma_{\rm SFR}$ relation is very steep with a slope $n\sim{}2$, the inferred (based on $\CO$ observations) $\Sigma_\H2-\Sigma_{\rm SFR}$ relation flattens significantly at $\Sigma_\H2\lesssim{}20$ $M_\odot$ pc$^{-2}$. This effect is already pronounced on kpc scales, but becomes very strong on scales of 250 pc. In other words, we  predict that the $\H2$ gas depletion time derived from $\CO$ emission would appear to decrease, not increase, with decreasing surface density at $\Sigma_\H2\lesssim{}20$ $M_\odot$ pc$^{-2}$ on sufficiently small scales. This effect is a consequence of the anti-correlation between $X_\CO$ and $\Sigma_\H2$ at low gas surface densities, caused by the presence of large amounts of $\CO$-dark molecular gas (\citealt{2010ApJ...716.1191W, 2011ApJ...731...25K, 2011MNRAS.418..664N}; paper I; \citealt{2011MNRAS.412.1686S, 2012MNRAS.tmp.2537N}). 

For $20\,M_\odot\,{\rm pc}^{-2} \lesssim{} \Sigma_\H2 \lesssim{} 100\,M_\odot\,{\rm pc}^{-2}$, on the other hand, the X-factor plays only a small role and the steep slope ($n\sim{}2$) is recovered. At such gas surface densities the diffuse emission in the star formation maps is also expected to be less of an obstacle \citep{2012ApJ...745..183R}. Hence, the good news is that a robust measurement of the slope of the $\Sigma_\H2-\Sigma_{\rm SFR}$ relation based on $\CO$ data will be possible in areas of moderately high gas and SFR surface density. At low $\Sigma_\H2$, however, and especially on sub-kpc scales, an accurate determination of the slope on the basis of $\CO$ observations will be difficult.

\subsection{Comparison with $z\sim{}2$ galaxies}
\label{sect:highZ}

Studies of the $\Sigma_\H2-\Sigma_{\rm SFR}$ relation that are based on samples of galaxies from both the local Universe and from high redshift indicate that the slope is close to (but not quite) linear over a many orders of magnitude of $\Sigma_\H2$ \citep{2010MNRAS.407.2091G}. In particular, if one selects only non-interacting galaxies, the slope of the $\Sigma_\H2-\Sigma_{\rm SFR}$ relation is approximatively $\sim{}1.1-1.2$. This, however, appears to be driven by a steepening at \emph{high surface densities}, because the slope measured on intermediate $\H2$ surface densities is much closer to unity \citep{2008AJ....136.2846B}.

In our analysis we focus on \emph{non-interacting galaxies}, that presumably form stars in an approximatively steady state. We do not include interacting or merging galaxies in our discussion, as such objects can be scattered off the $\Sigma_\H2-\Sigma_{\rm SFR}$ relation \citep{2010MNRAS.407.2091G, 2010ApJ...714L.118D}.

While there are several ways to explain a non-linear slope of the $\Sigma_\H2-\Sigma_{\rm SFR}$ relation, we demonstrate below that our star formation and $\CO$ emission model coupled with an actually \emph{linear} $\Sigma_\H2-\Sigma_{\rm SFR}$ relation does predict a slightly non-linear slope of the inferred ($\CO$-based) $\Sigma_\H2-\Sigma_{\rm SFR}$ relation at high $\Sigma_\H2$, consistent with observations. In essence, we argue that the observed super-linear slope can be understood as an X-factor effect.

We note that the following predictions rest on the assumption that there is no significant trend of the $\CO$ line ratios or the $\CO$ brightness temperature with gas surface density. Our predictions are specifically for the $J=1\rightarrow{}0$ rotational transition line of $\CO$, while galaxies at $z\sim{}1-2$ are typically observed in $\CO$ line emission resulting from $J=2\rightarrow{}1$ or $J=3\rightarrow{}2$ transitions. Systematic trends in the emission line ratios could thus modify the slope of the $\Sigma_\H2-\Sigma_{\rm SFR}$ relation. A systematic variation of the brightness temperature with surface density would have a similar effect. So far there is no clear observational evidence that either of these assumptions is violated in steady-state, normally star forming galaxies at $z\sim{}1-2$.

\begin{figure*}
\begin{tabular}{cc}
\includegraphics[width=80mm]{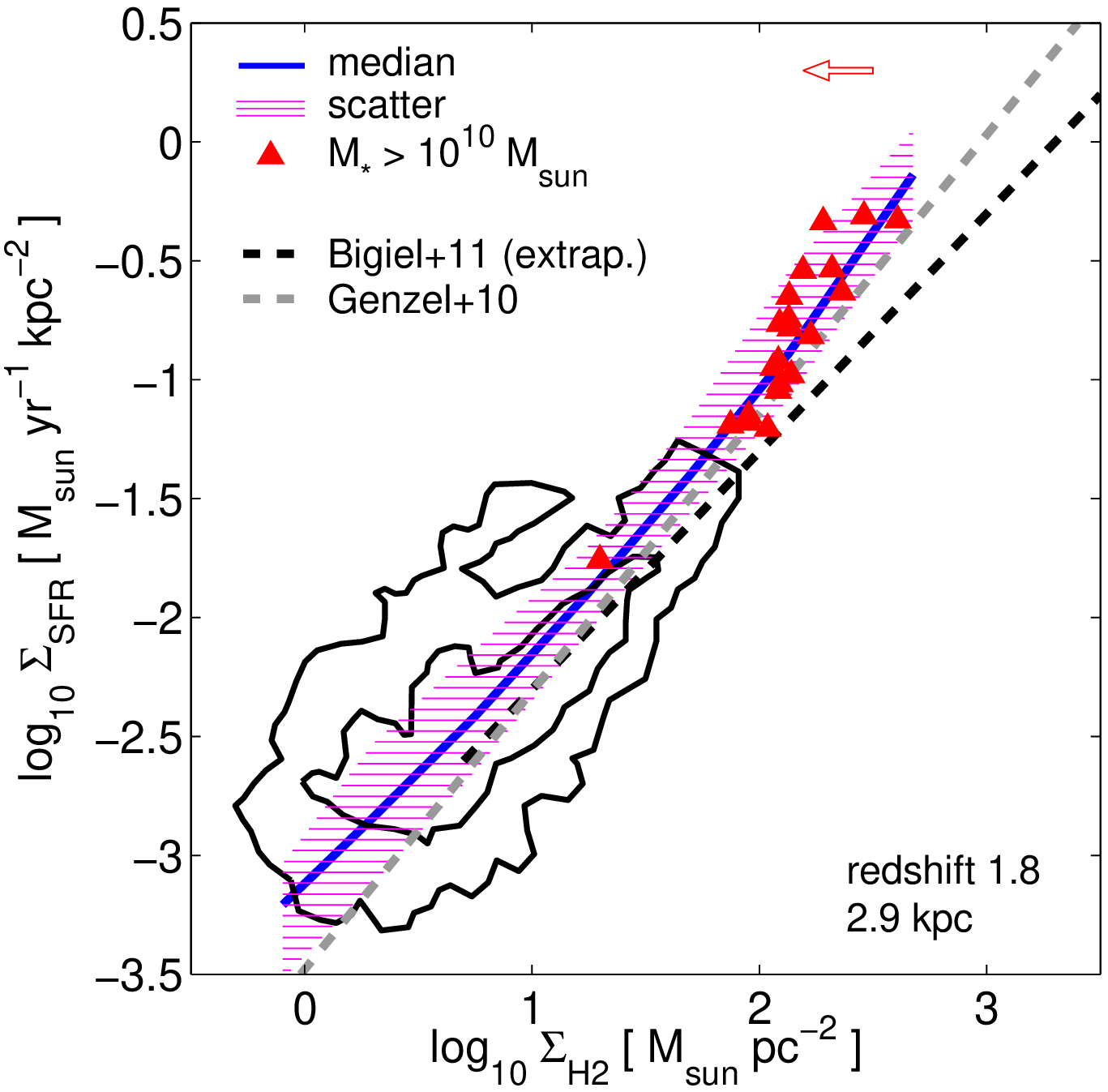} &
\includegraphics[width=80mm]{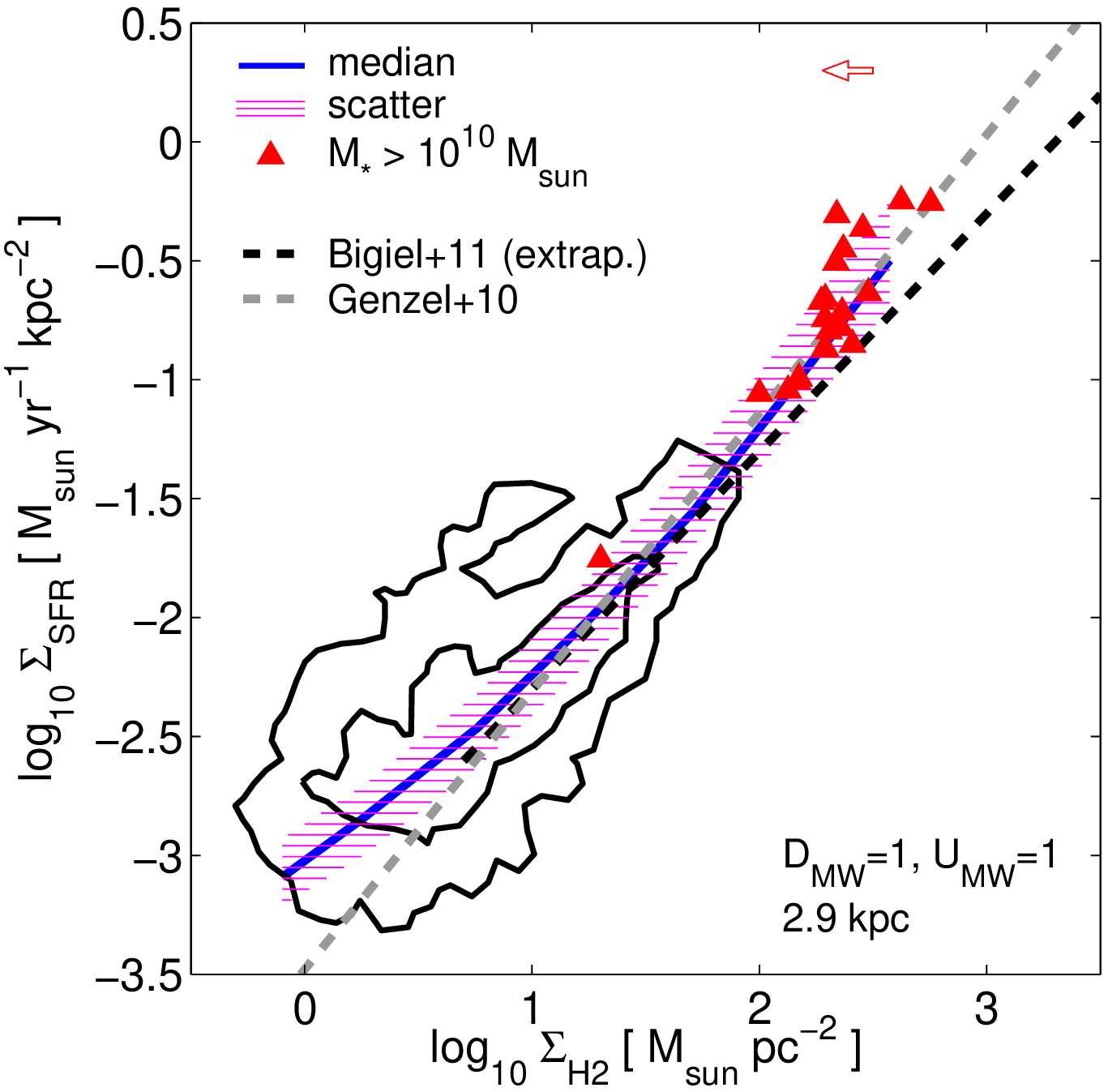} \\
\end{tabular}
\caption{The $\Sigma_\H2-\Sigma_{\rm SFR}$ relation as predicted by a fully cosmological, hydrodynamical simulation at $z\sim{}2$ (simulations HZ-csm, HZ-fid). Each panel shows the predictions of the ``inferred'' 
$\Sigma_\H2 - \Sigma_{\rm SFR}$ relation, i.e., the one where $\H2$ surface densities are derived from $^{12}\CO$ emission maps using a galactic conversion factor $X_{\CO,{\rm MW}}=2\times{}10^{20}$ K$^{-1}$ cm$^{-2}$ km$^{-1}$ s. (Left) the interstellar radiation fields and metallicities as computed self-consistently within the simulation at $z=1.8$. The $\H2$ weighted mean metallicity at this epoch is only  $\sim{}0.5\,Z_\odot$ but spans a broad range from (1-$\sigma$) $0.25$ to $0.75$ $Z_\odot$. This results in a value of $X_\CO$ that is above the galactic conversion factor and hence shifts the median of the inferred $\Sigma_\H2-\Sigma_{\rm SFR}$ relation slightly toward lower surface densities when compared with the observations by \cite{2011ApJ...730L..13B} for galaxies in the local Universe. The more interesting result is, however, that the slope of the $\Sigma_\H2-\Sigma_{\rm SFR}$ relation appears to steepen, in particular at high $\Sigma_{\rm SFR}$. (Right) The simulation is restarted at a slightly earlier epoch and continued for $\sim{}200$ Myr down to $z=1.8$, but this time with dust-to-gas ratios and UV radiation fields fixed to $D_\MW=1$ and $\UMW=1$, respectively. The steepening of the $\Sigma_\H2-\Sigma_{\rm SFR}$ relation remains visible and is therefore not a result of changes in the dust-to-gas ratios or interstellar radiation fields.
Symbols and lines are as in Fig.~\ref{fig:KS}b. In addition, the gray dashed line shows the fit to the observed $\Sigma_\H2-\Sigma_{\rm SFR}$ relation based on a large sample of low and high-z galaxies by \cite{2010MNRAS.407.2091G}. The red triangles mark the individual positions of simulated galaxies with stellar masses exceeding 10$^{10}$ $M_\odot$. All simulation predictions are based on our fiducial SF and CO model. The latter assumes a virial scaling of the CO line width. The red arrow at the top indicates the median shift in the inferred $\H2$ column density of the simulated galaxies if the CO line width would be fixed to a constant value of 3 km s$^{-1}$.
The figure shows that galaxies with high gas or SFR surface densities appear to deviate from a linear $\Sigma_\H2-\Sigma_{\rm SFR}$ relation, consistent with the observations of \cite{2010MNRAS.407.2091G}, despite the fact that the underlying relation between SFR and $\H2$ mass is perfectly linear. The super-linear slope ($\sim{}1.1-1.2$) of the inferred $\Sigma_\H2-\Sigma_{\rm SFR}$ relation is caused primarily by the increase of $X_\CO$ with increasing $\Sigma_\H2$ at high gas column densities. Our results apply only to galaxies that are in an equilibrium mode of star formation, not to starbursting galaxies. In the latter environments our CO model becomes unreliable and the galaxy-wide ratio between total molecular gas and the for SF relevant dense molecular gas ($n>10^4$ cm$^{-3}$) may change  \citep{2004ApJ...606..271G, 2012ApJ...745..190L, 2012arXiv1202.1803P}.
}
\label{fig:KSplot}
\end{figure*}

Our predictions for the inferred ($\CO$ based) $\Sigma_\H2-\Sigma_{\rm SFR}$ relation at $z\sim{}2$ are shown in Fig.~\ref{fig:KSplot}a. Galaxies with inferred molecular gas densities $\Sigma_\H2\gtrsim{}100$ $M_\odot$ pc$^{-2}$ are shifted off the actually linear $\Sigma_\H2-\Sigma_{\rm SFR}$ relation. The origin of this offset is the small, but systematic, variation of the X-factor with surface density. 

As discussed in paper I, at high $\Sigma_\H2$ the $\CO$ emission from a small ISM patch ($\sim{}20-100$ pc) ceases to scale linearly with the gas column density (and thus the $\H2$ surface density) due to the increased optical thickness at the line center of the $\CO$ emission line. This effect is somewhat, but not fully, compensated by an increase in the width of the emission line, so that overall there is a remaining increase of the X-factor with increasing $\H2$ surface density. For this reason, the use of a constant $\CO/\H2$ conversion factor leads to systematic shifts in the inferred $\Sigma_\H2-\Sigma_{\rm SFR}$ relation. At intermediate $\H2$ surface densities (the precise range depends on the spatial scale) the conversion factor is essentially constant and, hence, in this case the measured slope is predicted to be very close to linear, as observed \citep{2008AJ....136.2846B}.

Star forming galaxies at higher redshifts typically have lower $Z$ than galaxies of a similar mass in the local Universe, e.g., \cite{2008A&A...488..463M}. Our fully self-consistent cosmological simulation HZ-csm predicts that metallicities are only $\sim{}0.5\,Z_\odot$ at $z\sim{}2$. Hence, $D_\MW\sim{}0.5$ and one may wonder whether the combination of a high redshift, low metallicity sample and a low redshift, high metallicity sample could explain the observed super-linear relation. In order to test the importance of the metallicity dependence of $\CO/\H2$ conversion factor we rerun the HZ-csm simulation with fixed Milky-Way like ISM conditions (simulation HZ-fid, see \S\ref{sect:Sims}). 

Fig.~\ref{fig:KSplot}b shows that galaxies lie along the observed, slightly super-linear $\Sigma_\H2-\Sigma_{\rm SFR}$ relation, even if they had $D_\MW=1$. We therefore conclude that the super-linear slope is caused primarily by the scaling of $X_\CO$ with $\H2$ surface density, and that metallicity and dust-to-gas ratio variations have only a small effect.

We stress again that our $\CO$ emission model rests on a number of assumptions as pointed out above. If either of these assumptions were broken, an alternative explanation for the super-linearity of the $\Sigma_\H2-\Sigma_{\rm SFR}$ relation would be required. 

Systematic variations of the $\CO$ line ratios or the $\CO$ brightness temperature with gas surface density could be responsible for a difference between the intrinsic slope of the $\Sigma_\H2-\Sigma_{\rm SFR}$ relation and the slope derived from $\CO$ observations \citep{2011MNRAS.412..287N, 2012MNRAS.tmp.2537N}. Alternatively, a variation of the $\CO$ line ratios or brightness temperature with \emph{redshift} could lead to a systematic $z$ dependence of the normalization of the observed $\Sigma_\H2-\Sigma_{\rm SFR}$ relation. This can produce an artificial trend with surface density if a galaxy sample is used in which $\Sigma_\H2$ (and $\Sigma_{\rm SFR}$) strongly correlate with galaxy redshift. Finally, the $\Sigma_\H2-\Sigma_{\rm SFR}$ relation may actually get steeper at high column densities. For instance, it has been suggested that star formation becomes more efficient at high column densities because external pressure on molecular clouds shifts the balance between gravity and turbulent support \citep{2005ApJ...630..250K}. Upcoming observations with the Atacama Large Millimeter Array will hopefully enable us to distinguish between our model and these alternatives (e.g., see \citealt{2012arXiv1203.5280F}).

\subsection{The scatter in the $\Sigma_\H2-\Sigma_{\rm SFR}$ relation}
\label{sect:Scatter}

There are many effects and processes that could, in principle, contribute to the scatter in the $\Sigma_\H2-\Sigma_{\rm SFR}$ relation. Clearly, scatter can arise from (1) uncertainties related to the method of estimating SFRs, (2) uncertainties related to the estimation of $\H2$ masses and surface densities, (3) a possible non-linearity of the star formation process, and (4), any systematic uncertainties in the observables that were not accounted for. We will focus in this paper on the scatter sources (1) and (2). The potential role of (3) is discussed in detail in \cite{2011ApJ...732..115F}. We do not attempt to model sources that fall under category (4), since they are not intrinsic but depend on the specifics of the observational survey.\newline

\emph{SFR estimates}: The stellar mass that was formed over some past time interval will, in general, not coincide with the SFR that is expected based on the present $\H2$ mass. In other words, as shown in \cite{2011ApJ...732..115F}, the use of a \emph{time-averaged} SFR as an estimator of the \emph{ensemble-average} SFR introduces scatter. The following (not necessarily distinct) mechanisms fall under this category:
\begin{itemize}
\item \emph{discreteness of star formation}: star formation occurs in individual star formation events, i.e., is clustered in time,
\item \emph{stochasticity of star formation}: star formation relations on small scales hold  only in an (ensemble) average sense.
\item \emph{fluctuations in the $\H2$ abundance}: $\H2$ densities and, thus, the ensemble average SFRs may fluctuate on short time scales \citep{2007ApJ...659.1317G}, while the SFRs derived from tracers are smoothed because of the inherent time averaging,
\item \emph{evolutionary processes}: the conversion efficiency from gas to stars may change over the lifetime of molecular clouds
(\citealt{2011ApJ...729..133M}; cf. \citealt{2011ApJ...727L..12F}; see also \citealt{2010ApJ...722.1699S, 2010ApJ...722L.127O}),
\item \emph{imperfect tracers}: observational tracers of SFRs (e.g., $\Ha$ luminosities) may provide only approximate estimates of time-averaged SFRs because the proper conversion factor is not known exactly (e.g., it depends on the precise star formation history).
\end{itemize}

Observations indicate that most stars form in quantized units (embedded star clusters, \citealt{2003ARA&A..41...57L}) and hence that star formation is a discretized process. While this  inevitably makes star formation a stochastic process (in the above sense), it does not mean that all stochasticity in star formation arises from discreteness effects. For instance, the SFR in a molecular cloud depends on more than just its molecular mass. Magnetic fields, cosmic ray density, or the virialization state of the cloud will play a role to some extent. Marginalizing over these additional control parameters will give rise to an apparent stochasticity of star formation. In the formalism of \S\ref{sect:SFmodel}, the gas depletion time  $\tau_{\rm dep}$ becomes a function of these additional parameters and its replacement with some appropriately averaged, constant depletion time leads to the appearance of stochasticity. For instance, a star formation efficiency that evolves over the lifetime of molecular clouds can be interpreted in this sense. Here, the age of the cloud is the additional control parameter that determines $\tau_{\rm dep}$.

Our numerical models account for scatter caused by fluctuations in the $\H2$ abundance, the use of observational SFR tracers, the time discreteness of star formation, and any incidental stochasticity, but do not include the potential contributions from any additional control parameters. \newline

\emph{$\H2$ mass estimates}: As mentioned above, uncertainties in the estimates of the $\H2$ surface densities contribute to scatter in the $\Sigma_\H2-\Sigma_{\rm SFR}$ relation. Hence, variations in the $\CO/\H2$ conversion factor are a potential source of scatter for observational studies that are based on $\CO$ emission. As discussed in paper I, the X-factor can vary significantly, even for a fixed dust-to-gas ratio and $\H2$ column density, because $\Sigma_\H2$ depends on the product of $\H2$ mass fraction and total gas density, while the $\CO$ emission depends on the latter but not the former.\newline

\emph{Systematic variations of observables}: 
We measure the scatter under the condition that the dust-to-gas ratio and the interstellar radiation field are kept fixed (in the sense of \S\ref{sect:SubGridModeling}). This is an important point since the X-factor depends strongly on the former quantity (but only weakly on the latter, see paper I). Hence, variations in the dust-to-gas ratio lead to systematic modulations of conversion factor and, if unaccounted for, to scatter in the $\Sigma_\H2-\Sigma_{\rm SFR}$ relation. 

Often metallicity variations within a given galaxy are a strong function of galacto-centric radius \citep{1971ApJ...168..327S} and can be large from one galaxy to another. Hence, such systematic X-factor variation will appear as a galaxy-to-galaxy scatter \citep{2011AJ....142...37S}. Since the amount of scatter that is created in this way \emph{depends on the particular selection function} of the galaxy sample, we do not include galaxy-to-galaxy scatter in this analysis. Our predictions should therefore be compared with observations based on samples of galaxies with approximatively the same dust-to-gas ratio. \newline

\begin{figure*}
\begin{tabular}{cc}
\includegraphics[width=80mm]{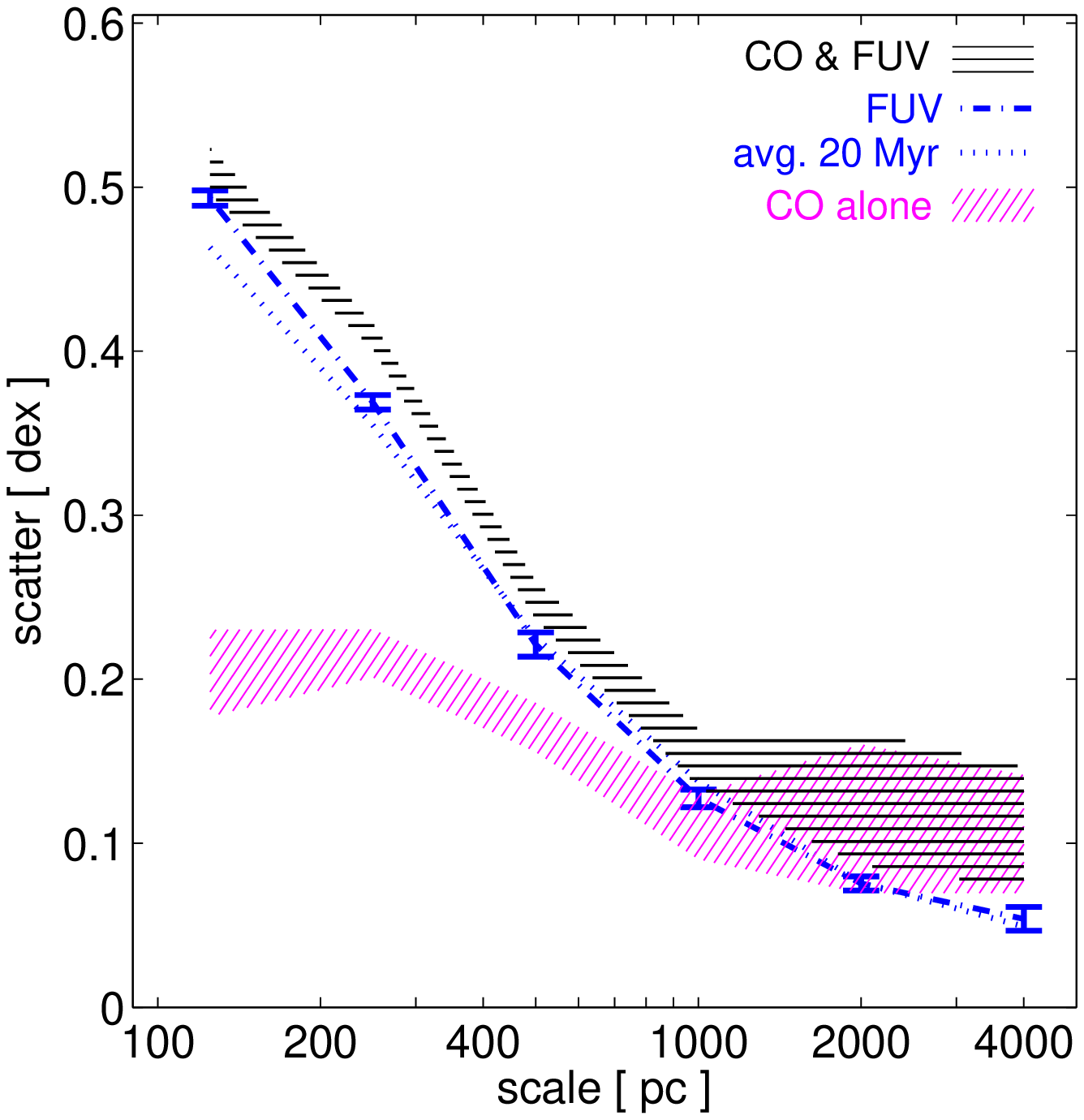} & 
\includegraphics[width=80mm]{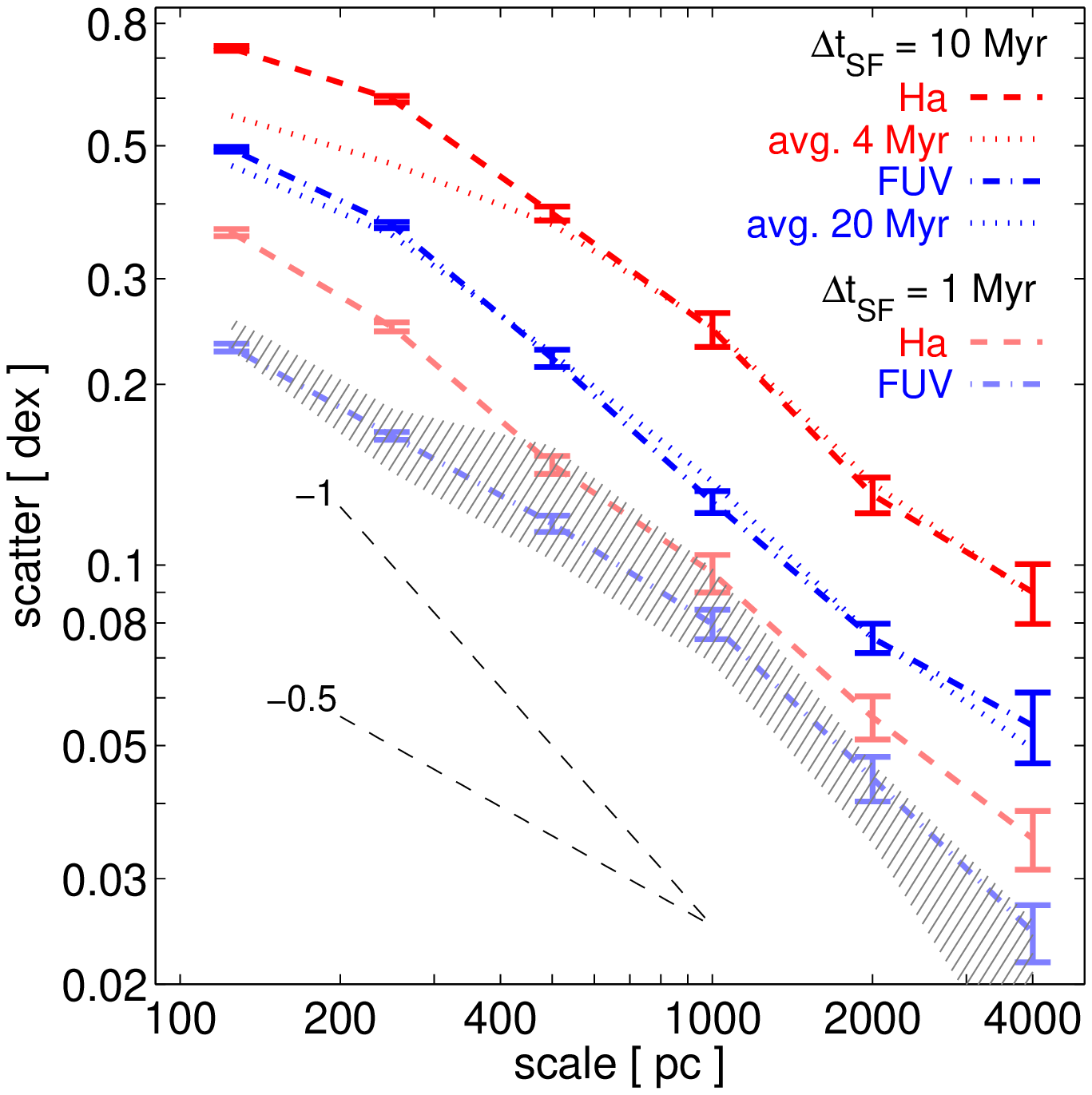} \\
\end{tabular}
\caption{Scatter in the  $\Sigma_\H2-\Sigma_{\rm SFR}$ relation as function of spatial averaging (resolution) scale for a galaxy with Milky-Way like ISM properties ($D_\MW=1$, $\UMW=1$). In both panels the scatter is measured over the range 10 $M_\odot$ pc$^{-2}  <  \Sigma_\H2 < 100$ $M_\odot$ pc$^{-2}$. Regions with $\Sigma_{\rm SFR}<3\times{}10^{-4}$ $M_\odot$ yr$^{-1}$ kpc$^{-2}$ or $I_\CO<0.2$ K km s$^{-1}$ are excluded from the analysis. (Left)  The blue dot-dashed line shows the scatter that arises when $\Sigma_\H2$ is know exactly, but SFR are inferred from FUV luminosities. The blue dotted lines marks the analogous result when SFR are derived from the stellar masses formed within the last 20 Myr. Scatter in $X_\CO$ at fixed CO emission leads by itself  to a scatter in the inferred $\Sigma_\H2-\Sigma_{\rm SFR}$ relation of the order of 0.1-0.2 dex and is shown as the magenta hashed region. The lower and upper boundaries of this region correspond to the cases of virial scaling of the CO line width vs constant line width, respectively (see text). Fluctuations in $X_\CO$ are not an important source of scatter on sub-kpc scales (at fixed dust-to-gas ratio and interstellar radiation field), but become increasingly relevant on scales of $\sim{}$kpc and above.
Finally, the black horizontally hashed region shows the combined scatter that takes into account both the scatter in $X_\CO$ and the scatter associated with the estimations of SFRs. (Right) This panel shows how the scatter depends on the SF tracer (FUV, $\Ha$, or simple time averaged SFR) and on assumptions about the stochasticity of the SF process (see legend). The critical parameter is the average time $\Delta{}t_\mathrm{SF}$ between SF events at a given site within the galaxy (see text). Specifically, the upper 4 lines show the scatter as derived from the various SF tracers for $\Delta{}t_\mathrm{SF} =10$ Myr (our fiducial value), while the two lines just below correspond to $\Delta{}t_\mathrm{SF} =1$ Myr. The gray hashed region shows the scatter for a run with $\Delta{}t_\mathrm{SF}=0.1$ Myr. This scatter results from the mismatch in time scales between SFRs that are averaged over the lifetime of a particular tracer (4 Myr - lower boundary; 20 Myr - upper boundary) and $\H2$ masses that are observed at a given instant. 
The figure shows that stochastic effects play a crucial role in determining the overall scatter in the $\Sigma_\H2-\Sigma_{\rm SFR}$ relation. Furthermore, modulo $X_\CO$ effects, the scatter decreases with scale $l$ roughly as a power law $\propto{}l^{-\alpha}$, with $\alpha\approx{}0.5-0.7$, consistent with the findings and interpretation given by \cite{2011ApJ...732..115F}.}
\label{fig:Scat}
\end{figure*}

In Fig.~\ref{fig:Scat}a we plot the scatter of the $\Sigma_\H2-\Sigma_{\rm SFR}$ relation as a function of spatial averaging scale and separate the contributions that result from the use of time-averaging tracers of star formation and uncertainties in $X_\CO$, respectively. The scatter is computed from all regions with an (inferred) $\H2$ column density between 10 and 100 $M_\odot$ pc$^{-2}$, a $\CO$ velocity integrated intensity equal to or larger than $0.2$ K km s$^{-1}$, and a minimum SFR surface density of $3\times{}10^{-4}$ $M_\odot$ yr$^{-1}$ kpc$^{-2}$.
These limits are chosen to roughly mimic typical values encountered in observational studies and it is clear that the exact numerical predictions will depend to some extent on these limits. In particular, the choice of the minimal $\Sigma_{\rm SFR}$ is crucial, since the scatter is measured in $\log{}\Sigma_\H2-\log{}\Sigma_{\rm SFR}$ space and, if not removed, regions with very low star formation would contribute enormously to the scatter (even worse, regions with zero star formation would make the scatter formally infinite).

Our numerical modeling predicts that X-factor variations induce a scatter of the order of $\sim{}0.1-0.2$ dex. Fig.~\ref{fig:Scat}a shows that this  scatter may be relevant on scales $\sim{}$kpc and above, but on smaller scales the total scatter is primarily due to the use of time averaged SFRs, at least for our fiducial choice $\Delta{}t_{\rm SF}=10$ Myr. Again we stress that this analysis assumes that variations in the dust-to-gas ratio and $\CO$ brightness temperature are small or accounted for. Fig.~\ref{fig:Scat}a also shows that the scatter that results from the use of time-averaged SFRs is a strong function of scale. It reaches $\sim{}0.5$ dex at $\sim{}100$ pc, but is only $\sim{}0.1$ dex at kpc scales.

Furthermore, there is little difference between the use of FUV luminosity-based SFRs and the use of the actual time-averaged SFR over the last 20 Myr. Hence, a varying FUV-to-stellar mass conversion factor (see \emph{imperfect tracers} above) contributes little to the scatter. This rules out the suggestion by \cite{2012arXiv1202.2873L} that the variation of the FUV luminosity over the lifetime of a single stellar population (SSP) dominates the scatter, at least on scales of $\sim{}100$ pc and above. In fact, if the luminosity-weighted SFRs differ little from the time-averaged SFRs, e.g., if SFRs are constant, then the luminosity evolution of the tracer becomes completely irrelevant. This can be seen from the following simple analysis. 

The total spectral luminosity from a Lagrangian volume element at time $t$ is given as
\[
L_\nu(t) = \int_{-\infty}^{t} {\rm SFR}(t') \phi_\nu(t-t') dt'.
\]
Here, $\phi_\nu(t)$ is the spectral luminosity from a SSP of unit mass and age $t$. This can be rewritten as 
\begin{equation}
L_\nu(t) = \langle{} {\rm SFR} \rangle{}_{\phi_\nu}(t) E_\nu,
\end{equation}
where
\begin{align*}
E_\nu &= \int_0^{\infty}\phi_\nu(t')dt',\,{\rm and}\\
\langle{} {\rm SFR} \rangle{}_{\phi_\nu}(t)&=\frac{1}{E_\nu}\int_0^{\infty} {\rm SFR}(t-t')\phi_\nu(t')dt'
\end{align*}
are the total spectral energy emitted by an SSP of unit mass and the luminosity weighted SFR, respectively. Hence, if time-averaged and luminosity weighted SFRs trace each other closely, then the constant $E_\nu$ is the perfect (i.e., scatter-free) conversion factor between tracer luminosity and time-averaged SFR. The validity of this statement does not depend on the form of $\phi_\nu$. 

We show in Fig.~\ref{fig:Scat}b that the scatter in the $\Sigma_\H2-\Sigma_{\rm SFR}$ relation depends on both the star formation tracer and the time discreteness parameter $\Delta{}t_{\rm SF}$. We find that $\Ha$-based SFR estimates lead to more scatter in the $\Sigma_\H2-\Sigma_{\rm SFR}$ relation than the use of FUV flux as a tracer. Hence, the star formation tracer with the shorter lifetime ($\Ha$) leads to larger scatter. Similarly, when we estimate the SFR based on the actual stellar mass formed within the past 4 Myr and the 20 Myr, we find that the use of a shorter averaging time leads to more scatter in the $\Sigma_\H2-\Sigma_{\rm SFR}$ relation.
 
The averaging timescales of 4 and 20 Myr  correspond roughly to the luminosity weighted timescales of $\Ha$ and FUV emission \citep{2012arXiv1202.2873L}. It is therefore not entirely surprising that the scatter predictions computed using these time-averaged SFRs are similar to the predictions that use $\Ha$ and FUV based tracers. 
It demonstrates that $\Ha$ and FUV based tracers can, to a good degree of approximation, be treated as a top-hat filter with a width of $\sim{}4$ Myr and $\sim{}20$ Myr, respectively. This correspondence will break, however, if the scales are small enough and the lifetimes of the particular tracer short enough such that luminosity weighted SFRs and time-averaged SFRs begin to differ substantially. For  $\Ha$ based SFR estimates this appears to happen on scales of $<400$ pc, while for FUV based tracers the effect is small even on spatial averaging scales of $\sim{}100$ pc.

Fig.~\ref{fig:Scat}b also shows that the scatter depends on the average time between star formation events $\Delta{}t_{\rm SF}$. As expected a shorter $\Delta{}t_{\rm SF}$ means that individual star formation events involve less stellar mass but occur at a higher rate, which reduces the scatter. The sharp drop in the scatter when $\Delta{}t_{\rm SF}$ is reduced to 1 Myr proves that much of the scatter in our fiducial $\Delta{}t_{\rm SF}=10$ Myr model arises from a single source, the discreteness of star formation. Hence, observational estimates of the scatter can be used to put tight constraints on the value of $\Delta{}t_{\rm SF}$. In addition, a systematic observational study of the scatter in the  $\Sigma_\H2-\Sigma_{\rm SFR}$ relation as function of ISM environment would allow to determine whether (and how) $\Delta{}t_{\rm SF}$ depends on ISM properties.

In order to assess how much of the total scatter is caused by time variations in the $\H2$ abundance (and not related to time discreteness of star formation), we also show the scatter for a run with $\Delta{}t_{\rm SF}=0.1$ Myr. This timescale is close to the smallest dynamical time step in our simulation and, hence, effectively eliminates any discreteness (beyond that dictated by the simulation time step) in the star formation model. We then measure the scatter in the true $\Sigma_\H2-\Sigma_{\rm SFR}$ relation using time-averaged SFRs (using 4 Myr and 20 Myr as averaging times). In this way we isolate the scatter that is caused by the observational actuality that SFRs are time-averaged quantities, but $\H2$ masses are observed at a particular instant. The scatter that results in this way is relatively small, see Fig.~\ref{fig:Scat}b. On super-kpc scales  it is dominated by the scatter that is caused by $X_\CO$ fluctuations and on  sub-kpc scales by the scatter due to the discreteness of star formation. However, we point out that time variations in the $\H2$ abundance couple in a non-linear way to the Poisson noise of individual star formation events (see appendix). Hence, they contribute to the scatter caused by the discreteness of star formation and, hence, cannot be neglected.

In \cite{2011ApJ...732..115F} we found that noise inserted by hand on small scales leads to scatter in the $\Sigma_\H2-\Sigma_{\rm SFR}$ relation that decreases roughly as power law $\propto{}l^{-\alpha}$, where $\alpha\approx{}0.5$, with increasing spatial averaging scale $l$. Fig.~\ref{fig:Scat}b shows that the scatter due to the stochastic nature of star formation follows this scaling approximatively. In particular, it is clearly less steep than a $\alpha=1$ scaling which would be the naive expectation if the gas were arranged in disk of uniform density. As discussed in \cite{2011ApJ...732..115F} the scaling deviates from $\alpha=1$ because the density distribution of the ISM, determined by turbulence, is far from being uniform.

How does the scatter that is predicted for our fiducial choice $\Delta{}t_{\rm SF}=10$ Myr compare with observations? Such a comparison is not straightforward since observational estimates of the scatter depend on choices in the methodology. For instance, the treatment of diffuse emission does not only affect the inferred slope of the $\Sigma_\H2-\Sigma_{\rm SFR}$ relation, but also the scatter. Furthermore, the measured scatter depends on the surface density range of the fit. With these caveats in mind we will now compare our predictions to observational studies that give quantitive estimates of the scatter at a given scale.

\cite{2011ApJ...730...72R} infer a scatter of about 0.3-0.4 dex for SFRs based on $\Ha$ luminosities on $\sim{}0.5$ kpc scales. The scatter is lower ($\sim{}0.1-0.3$ dex) when they use tracers with longer lifetimes (FUV + 24$\mu{}m$). Our model predicts a scatter of $\sim{}0.4$ dex for $\Ha$ based SFRs and a scatter of $\sim{}0.2$ dex if FUV luminosities are used to trace star formation  on $\sim{}0.5$ kpc scales. This quantitative agreement is a further justification of the choice $\Delta{}t_{\rm SF}=10$ Myr.

\cite{2010A&A...510A..64V} investigate how the scatter increases with increasing spatial resolution. Their Fig. 4 shows a scatter of $\sim{}0.4$ dex on  360 pc scales, $\sim{}0.35$ dex on 720 pc scales, and $\sim{}0.3$ dex on 1.4 kpc scales. This is a somewhat shallower scaling than our predictions ($\sim{}0.4$ dex on 500 pc scales, $\sim{}0.25$ dex on kpc scales and 0.1-0.15 dex on super-kpc scales), see Fig.~\ref{fig:Scat}b. However, their scatter is computed by an iterative clipping method that would tend to underestimate the scatter if the scatter is large, i.e., on smaller scales.

\cite{2010ApJ...722.1699S} compare the gas depletion times $\tau_\CO$ and $\tau_\H2$ in apertures centered on peaks of $\CO$ and $\Ha$ emission, respectively. Unfortunately, they do not report the scatter in the $\Sigma_\H2-\Sigma_{\rm SFR}$ relation. However, we can compare the change of $\log_{10}(\tau_\CO-\tau_\H2)$, a crude estimator of the scatter, with changing spatial averaging scale using their Fig. 3. This results in a scaling similar to the predictions given in our Fig.~\ref{fig:Scat}.

To summarize, we find that for galaxies with Milky-Way like ISM conditions most of the scatter in the $\Sigma_\H2-\Sigma_{\rm SFR}$ relation is a consequence of the time discreteness of star formation. Systematic variations in the SFR tracer conversion factors are only relevant for tracers with short lifetimes (e.g., $\Ha$) \emph{and} when observations are done on sufficiently small scales ($<400$ pc). Variations of the $\H2/\CO$ conversion factor can dominate the scatter on kpc scales and above, but are unimportant on much smaller scales. Finally, fluctuations in the $\H2$ abundance play a supporting role, enhancing the scatter caused by the discreteness of the star formation process.

\section{Discussion}
\label{sect:DiscussionScatter}

How does the star formation model presented in this paper relate to alternative interpretations of the scatter in the  $\Sigma_\H2-\Sigma_{\rm SFR}$ relation? 

A commonly made suggestion is that the relation ``breaks down'' on small scales \citep{2010ApJ...721..383M, 2010ApJ...722L.127O, 2010ApJ...722.1699S} because molecular clouds pass through evolutionary phases in which they transform from $\CO$-bright, but star-less, clouds to star forming regions with little surrounding molecular gas. One of the difficulties with a picture in which most of the molecular gas in a galaxy goes through a well defined sequence of stages is the following. It does not explain why the $\H2$ depletion time of molecular clouds with embedded young stellar objects is only a few 100 Myr \citep{2010ApJ...724..687L}, while the depletion time on galactic scales is an order of magnitude larger \citep{2011ApJ...730L..13B}. In fact, this observation, originally used as evidence to support long lifetimes of molecular clouds \citep{1974ApJ...192L.149Z}, implies that  the majority of the molecular gas in the galaxy has to be in a non-star forming state \citep{2000ApJ...530..277E}, possibly in either non-star forming clouds that will be dispersed before star formation has a chance to begin, or in unbound molecular associations. 

The way our model addresses this problem is that the observed gas depletion time of a molecular \emph{and} star forming region is smaller than the average depletion $\tau_{\rm dep}$ by a factor $\sim{}\Delta{}t_*/\Delta{}t_{\rm SF}$ (see appendix). Observations of individual molecular clouds often derive SFRs based on counts of young stellar objects ($\Delta{}t_*=1-2$ Myr) or by using tracers of ionizing radiation from massive stars ($\Delta{}t_*\sim{}$ few Myr), e.g., $\Ha$ emission or free-free radio emission. Consequently, $\Delta{}t_*<\Delta{}t_{\rm SF}=10$ Myr and the observed gas depletion time in star forming regions is shorter than $\tau_{\rm dep}$. This is balanced by the large (formally infinite) depletion time in molecular regions that are not currently star forming.

It has also been suggested that the scatter in the $\Sigma_\H2-\Sigma_{\rm SFR}$ relation is caused by variations in the ratio between molecular gas and \emph{dense} ($\gtrsim{}10^4$ cm$^{-3}$) molecular gas \citep{2012ApJ...745..190L}. Since our model does not explicitly follow dense gas, such variations could be included as an additional stochastic component in the star formation model, see \S\ref{sect:Scatter}. However, the tight correlation between dense gas and star formation in many different environments \citep{2012ApJ...745..190L, 2012arXiv1202.1803P} is suggestive of an alternative way. In the context of our model the only thing required is to re-interpret the word ``individual star formation event'' as ``individual dense gas formation event'' and to assume that, once gas becomes very dense, star formation is inevitable and will proceed on the $\sim{}$free-fall time of the respective dense gas clump. This modification of our model does not specify the physical mechanism for the sudden increase in gas density, but several plausible options exist, e.g., cloud collisions \citep{2009ApJ...700..358T}.

This re-interpretation of our star formation model accounts, by construction, for the observed tight correlation between dense gas and star formation rate. Furthermore, the ratio between dense (i.e., star forming) molecular gas and all molecular gas in a given region will vary depending on how many ``dense gas formation events'' have occurred in the region. In this model both the scatter in the $\Sigma_\H2-\Sigma_{\rm SFR}$ relation and the scatter in the mass ratio between dense and not-so-dense gas on small scales are a consequence of the discrete formation of dense gas clumps out of molecular gas.

Further potential contributors to the scatter include the incomplete sampling of the IMF and the drifting of stars out of their parent molecular clouds. These scatter sources are unlikely to be relevant given the relatively large spatial scales ($\gtrsim{}100$ pc) and SFR surface densities ($>10^{-3}$ $M_\odot$ yr$^{-1}$ kpc$^{-2}$) in our study (see \citealt{2010ApJ...722L.127O}).

We conclude that in the context of our star formation model there is no ``break down'' of the $\Sigma_\H2-\Sigma_{\rm SFR}$ scaling relation. Instead, the proper interpretation is that the discrete and stochastic nature of star formation becomes evident as observations probe smaller and smaller scales. 

\section{Summary and Conclusions} 
\label{sect:Conclusions}

In this paper we studied the slope and the scatter of the $\Sigma_\H2-\Sigma_{\rm SFR}$ relation using cosmological galaxy formation simulations coupled with models for star formation, $\H2$ chemistry,  and $\CO$ emission. Our focus is especially on the role of the $\CO/\H2$ conversion factor. We found that X-factor variations with surface density can result in significant biases of the measured slope. In particular, at high spatial resolution (few 100 pc or better) and sufficiently low surface densities ($\Sigma_\H2<20\,M_\odot\,{\rm pc}^{-2}$) the slope inferred from $\CO$ observations is shallower than the actual slope. In contrast, the inferred slope becomes steeper than the true slope if galaxies with high $\H2$ surface densities (above $100$ $M_\odot$ pc$^{-2}$) are included in the sample, providing an possible explanation for the slightly super-linear slope of the $\Sigma_\H2-\Sigma_{\rm SFR}$ relation seen at high gas surface densities (e.g., \citealt{2010MNRAS.407.2091G}). Yet, we also showed that measurements at $\gtrsim{}500$ pc resolution over a surface density range often studied in samples of nearby galaxies, $10\,M_\odot\,{\rm pc}^{-2}<\Sigma_\H2<100\,M_\odot\,{\rm pc}^{-2}$, are essentially unbiased.

Variations in the X-factor contribute to the scatter in the  $\Sigma_\H2-\Sigma_{\rm SFR}$ relation (of the order of $\sim{}0.1-0.2$ dex), dominating over many other scatter sources when the spatial resolution of the survey is $\sim{}$kpc or larger. This even holds if there are no significant spatial variations of the dust-to-gas ratio or the interstellar radiation field. Such variations are expected in a heterogeneous sample of galaxies, leading to additional galaxy-to-galaxy scatter with an amount that depends on the properties of the particular galaxy sample (e.g., \citealt{2011AJ....142...37S}). On sub-kpc scales, however, spatial variations in the $\CO/\H2$ conversion factor contribute little to the overall scatter (assuming a fixed dust-to-gas ratio). On such scales much of the scatter is a consequence of the fact that the measured, time-averaged SFRs differ from the SFRs that are expected based on the present amount of $\H2$. 

We demonstrated that the scatter in the  $\Sigma_\H2-\Sigma_{\rm SFR}$ relation (on scales of $\sim{}100$ pc and larger) is primarily a consequence of the discreteness of the star formation process. The luminosity evolution of SFR tracers can become relevant for tracer with short lifetimes, e.g., $\Ha$, on small scales (less then a few 100 pc). For FUV-based SFRs, however, the scatter does not differ significantly from the scatter based on a hypothetical tracer with a plain 20 Myr lifetime. The differences in the timescales between SFR tracers (at least a few Myr, up to 100 Myr) and that of $\H2$ masses (essentially instantaneous measurements) does lead to some scatter, but it is typically dominated by scatter that results from X-factor variations (on super-kpc scales) and by scatter due to the discreteness of star formation (on sub-kpc scales).

The predictions made in the paper suggest a number of observational tests that could be used to constrain the presented numerical models. Fortunately, most of the more obvious tests, e.g., looking for a change in slope of the $\Sigma_\H2-\Sigma_{\rm SFR}$ relation at very low and very high surface densities (Fig~\ref{fig:KS2}b, Fig~\ref{fig:KSplot}), should be feasible with future ALMA observation. A clear test of the predictions of the Poisson star formation model should be possible with a systematic, observational study of how the scatter in the $\Sigma_\H2-\Sigma_{\rm SFR}$ relation scales with spatial scale and how it depends on the lifetimes of star formation tracers. Such a study would allow to constrain the discreteness of star formation and, more generally, would be a crucial guide for the development of the theoretical underpinnings of star formation in a galactic context.

\acknowledgements 

This work was supported in part by the DOE at Fermilab, by the NSF grant AST-0708154, by the NASA grant NNX-09AJ54G, and by the Kavli Institute for Cosmological Physics at the University of Chicago through the NSF grant PHY-0551142 and PHY-1125897 and an endowment from the Kavli Foundation. The simulations used in this work have been performed on the Joint Fermilab - KICP Supercomputing Cluster, supported by grants from Fermilab, Kavli Institute for Cosmological Physics, and the University of Chicago. This work made extensive use of the NASA Astrophysics Data System and {\tt arXiv.org} preprint server.

\appendix
\label{sect:app}

In this appendix we give a more formal definition of the Poisson star formation model introduced in \S\ref{sect:SFmodel} and describe its implementation in the ART code. The model assumes that the number of individual star formation events $N_{\Delta{}t}$ in a given time interval $\Delta{}t$ is a Poisson random variable with a mean and variance $\lambda=\frac{\Delta{}t}{\Delta{}t_{\rm SF}}$. Here, $\Delta{}t_{\rm SF}$ is the average time interval between two star formation events. The continuum, non-stochastic limit corresponds to $\Delta{}t_{\rm SF}\rightarrow{}0$. 

The ensemble average SFR density in a given region is proportional to the $\H2$ density, i.e.,
\begin{equation}
\langle{}\dot{\rho}_*\rangle{}(t) = \frac{\rho_\H2}{\tau_{\rm dep}},
\end{equation}
where the gas depletion time $\tau_{\rm dep}$ is assumed to be a constant. The ensemble average SFR in a given fixed volume $l_{\rm SF}^3$ is simply $\mathcal{S} = \langle{}\dot{\rho}_*\rangle{}l_{\rm SF}^3$. We denote the time average of $\mathcal{S}$ over some interval $\Delta{}t$ as $\overline{\mathcal{S}}_{\Delta{}t}$.

The actual SFR, $S_{\Delta{}t}$, that occurred  in the volume $l_{\rm SF}^3$ during the time interval $\Delta{}t$ is a \emph{random variable} and defined as
\begin{equation}
\label{eq:SDt}
S_{\Delta{}t} = N_{\Delta{}t} \, \frac{\Delta{}t_{\rm SF}}{\Delta{}t}\, \overline{\mathcal{S}}_{\Delta{}t},
\end{equation}
which ensures that the ensemble average of $S_{\Delta{}t}$ equals $\overline{\mathcal{S}}_{\Delta{}t}$.
This equations shows that the stellar mass formed during an individual star formation event is simply given by $\overline{\mathcal{S}}_{\Delta{}t} \Delta{}t_{\rm SF}$, i.e., it scales linearly with the average SFR. 
 
The time interval $\Delta{}t$ may be some small, arbitrarily chosen time step. We are typically interested in measuring the mean and variance of the SFR (or of its logarithm) over some physical time interval $\Delta{}t_*\gg{}\Delta{}t$, e.g., over the lifetimes of star formation tracers such as $H\alpha$ or FUV emission. The actual SFR over $\Delta{}t_*$ is given by
\begin{equation}
S_{\Delta{}t_*} = \frac{\Delta{}t}{\Delta{}t_*}\sum_i S_{\Delta{}t,i} =  \frac{\Delta{}t_{\rm SF}}{\Delta{}t_*}\sum_i N_{\Delta{}t,i} \, \overline{\mathcal{S}}_{\Delta{}t,i},
\label{eq:SDts}
\end{equation}
with indices $i$ running over the $\Delta{}t_*/\Delta{}t$ time intervals of length $\Delta{}t$. 

It is worthwhile to have a closer look at (\ref{eq:SDts}). First, the expression on the right hand side implies that $S_{\Delta{}t_*}$  does not depend on the time step $\Delta{}t$ provided $\Delta{}t$ is sufficiently short compared with the typical time over which $\mathcal{S}$ fluctuates. Moments of $S_{\Delta{}t_*}$ and  $\log_{10}S_{\Delta{}t_*}$ depend on the timescales $\Delta{}t_{\rm SF}$ and $\Delta{}t_*$  only via the ratio $\Delta{}t_*/\Delta{}t_{\rm SF}$. The equation also allows us to estimate the $\H2$ depletion time $\rho_\H2\,l_{\rm SF}^3/S_{\Delta{}t_*}$ that an observer would measure in a star forming region ($N_{\Delta{}t_*}=\sum_i N_{\Delta{}t,i}\geq{}1$). If we assume that the $\H2$ density in the region remains constant during $\Delta{}t_*$, then the depletion time that the observer would infer is $\tau_{\rm dep}(\Delta{}t_*/\Delta{}t_{\rm SF})/N_{\Delta{}t_*}$, which is smaller than $\tau_{\rm dep}$ if $\Delta{}t_*<\Delta{}t_{\rm SF}$.

Equation (\ref{eq:SDts}) further shows that $S_{\Delta{}t_*}$ depends in a non-linear way on both the time evolution of the $\H2$ mass in a given region and the number of individual star formation events. Hence, we expect that the scatter in the $\Sigma_\H2-\Sigma_{\rm SFR}$ relation that arises from the discreteness of star formation depends in a non-linear way on both the time variations in the $\H2$ mass and on Poisson shot noise in the number of individual star formation events. 
To be quantitative let us approximate the scatter in the $\Sigma_\H2-\Sigma_{\rm SFR}$ relation with $\sigma_{\log_{10}S_{\Delta{}t_*}}$, the standard deviation of the logarithm of $S_{\Delta{}t_*}>0$. Let us further assume that $S_{\Delta{}t,i}$ fluctuates over time $\Delta{}t_*$ as a log-normal random variable, i.e., $S_{\Delta{}t,i}\propto{}e^{\sigma{}X}$, where $X$ is a Gaussian random variable with mean 0 and variance 1. The choice $\Delta{}t_*=20$ Myr, $\Delta{}t_{\rm SF}=10$ Myr and $\sigma=1.4$ results in $\sigma_{\log_{10}S_{\Delta{}t_*}}\approx{}0.60$. With the same set of parameters but $\Delta{}t_{\rm SF}\rightarrow{}0$ we find $\sigma_{\log_{10}S_{\Delta{}t_*}}\approx{}0.07$. If we assume $\sigma=0$, i.e., $\overline{\mathcal{S}}_{\Delta{}t,i}$ constant, we obtain $\sigma_{\log_{10}S_{\Delta{}t_*}}\approx{}0.24$. Hence, the variance of $\log_{10}S_{\Delta{}t_*}$ is not simply the sum of the variances caused by Poisson noise and $\H2$ fluctuations, respectively, but is determined to a large extent by their covariance. Using this simple test setup we also find that the scaling of $\sigma_{\log_{10}S_{\Delta{}t_*}}$ with  $\Delta{}t_*/\Delta{}t_{\rm SF}$ depends on the value of $\sigma$, although, for sufficiently large $\Delta{}t_*/\Delta{}t_{\rm SF}$, the scatter decreases with increasing $\Delta{}t_*/\Delta{}t_{\rm SF}$.
 
Our simulations adopt the Poisson star formation model in the following way. Every $\Delta{}t=10^5$ yr the code computes the ensemble average SFR $\mathcal{S}$ in each resolution element ($l_{\rm SF}\sim{}100$ pc, see Table~\ref{tab:sim}) based on the present $\H2$ mass. Then, a random realization of $N_{\Delta{}t}$ is drawn from a Poisson distribution with the mean $\Delta{}t/\Delta{}t_{\rm SF}$ and, if $N_{\Delta{}t}>0$, the code creates a stellar particle of mass $N_{\Delta{}t}\,\mathcal{S} \Delta{}t_{\rm SF} $.

\bibliographystyle{apj}

\begin{thebibliography}{100}
\expandafter\ifx\csname natexlab\endcsname\relax\def\natexlab#1{#1}\fi

\bibitem[{{Allende Prieto} {et~al.}(2001){Allende Prieto}, {Lambert}, \&
  {Asplund}}]{2001ApJ...556L..63A}
{Allende Prieto}, C., {Lambert}, D.~L., \& {Asplund}, M. 2001, \apjl, 556, L63

\bibitem[{{Asplund} {et~al.}(2004){Asplund}, {Grevesse}, {Sauval}, {Allende
  Prieto}, \& {Kiselman}}]{2004A&A...417..751A}
{Asplund}, M., {Grevesse}, N., {Sauval}, A.~J., {Allende Prieto}, C., \&
  {Kiselman}, D. 2004, \aap, 417, 751

\bibitem[{{Asplund} {et~al.}(2009){Asplund}, {Grevesse}, {Sauval}, \&
  {Scott}}]{2009ARA&A..47..481A}
{Asplund}, M., {Grevesse}, N., {Sauval}, A.~J., \& {Scott}, P. 2009, \araa, 47,
  481

\bibitem[{{Bertschinger}(2001)}]{2001ApJS..137....1B}
{Bertschinger}, E. 2001, \apjs, 137, 1

\bibitem[{{Bigiel} {et~al.}(2008){Bigiel}, {Leroy}, {Walter}, {Brinks}, {de
  Blok}, {Madore}, \& {Thornley}}]{2008AJ....136.2846B}
{Bigiel}, F., {Leroy}, A., {Walter}, F., {Brinks}, E., {de Blok}, W.~J.~G.,
  {Madore}, B., \& {Thornley}, M.~D. 2008, \aj, 136, 2846

\bibitem[{{Bigiel} {et~al.}(2011){Bigiel}, {Leroy}, {Walter}, {Brinks}, {de
  Blok}, {Kramer}, {Rix}, {Schruba}, {Schuster}, {Usero}, \&
  {Wiesemeyer}}]{2011ApJ...730L..13B}
{Bigiel}, F., {Leroy}, A.~K., {Walter}, F., {Brinks}, E., {de Blok}, W.~J.~G.,
  {Kramer}, C., {Rix}, H.~W., {Schruba}, A., {Schuster}, K.-F., {Usero}, A., \&
  {Wiesemeyer}, H.~W. 2011, \apjl, 730, L13

\bibitem[{{Blanc} {et~al.}(2009){Blanc}, {Heiderman}, {Gebhardt}, {Evans}, \&
  {Adams}}]{2009ApJ...704..842B}
{Blanc}, G.~A., {Heiderman}, A., {Gebhardt}, K., {Evans}, II, N.~J., \&
  {Adams}, J. 2009, \apj, 704, 842

\bibitem[{{Bolatto} {et~al.}(2011){Bolatto}, {Leroy}, {Jameson}, {Ostriker},
  {Gordon}, {Lawton}, {Stanimirovi{\'c}}, {Israel}, {Madden}, {Hony},
  {Sandstrom}, {Bot}, {Rubio}, {Winkler}, {Roman-Duval}, {van Loon},
  {Oliveira}, \& {Indebetouw}}]{2011ApJ...741...12B}
{Bolatto}, A.~D., {Leroy}, A.~K., {Jameson}, K., {Ostriker}, E., {Gordon}, K.,
  {Lawton}, B., {Stanimirovi{\'c}}, S., {Israel}, F.~P., {Madden}, S.~C.,
  {Hony}, S., {Sandstrom}, K.~M., {Bot}, C., {Rubio}, M., {Winkler}, P.~F.,
  {Roman-Duval}, J., {van Loon}, J.~T., {Oliveira}, J.~M., \& {Indebetouw}, R.
  2011, \apj, 741, 12

\bibitem[{{Bournaud} {et~al.}(2011){Bournaud}, {Chapon}, {Teyssier}, {Powell},
  {Elmegreen}, {Elmegreen}, {Duc}, {Contini}, {Epinat}, \&
  {Shapiro}}]{2011ApJ...730....4B}
{Bournaud}, F., {Chapon}, D., {Teyssier}, R., {Powell}, L.~C., {Elmegreen},
  B.~G., {Elmegreen}, D.~M., {Duc}, P.-A., {Contini}, T., {Epinat}, B., \&
  {Shapiro}, K.~L. 2011, \apj, 730, 4

\bibitem[{{Braine} {et~al.}(2001){Braine}, {Duc}, {Lisenfeld}, {Charmandaris},
  {Vallejo}, {Leon}, \& {Brinks}}]{2001A&A...378...51B}
{Braine}, J., {Duc}, P.-A., {Lisenfeld}, U., {Charmandaris}, V., {Vallejo}, O.,
  {Leon}, S., \& {Brinks}, E. 2001, \aap, 378, 51

\bibitem[{{Brown} \& {Vanden Bout}(1991)}]{1991AJ....102.1956B}
{Brown}, R.~L., \& {Vanden Bout}, P.~A. 1991, \aj, 102, 1956

\bibitem[{{Daddi} {et~al.}(2010{\natexlab{a}}){Daddi}, {Bournaud}, {Walter},
  {Dannerbauer}, {Carilli}, {Dickinson}, {Elbaz}, {Morrison}, {Riechers},
  {Onodera}, {Salmi}, {Krips}, \& {Stern}}]{2010ApJ...713..686D}
{Daddi}, E., {Bournaud}, F., {Walter}, F., {Dannerbauer}, H., {Carilli}, C.~L.,
  {Dickinson}, M., {Elbaz}, D., {Morrison}, G.~E., {Riechers}, D., {Onodera},
  M., {Salmi}, F., {Krips}, M., \& {Stern}, D. 2010{\natexlab{a}}, \apj, 713,
  686

\bibitem[{{Daddi} {et~al.}(2010{\natexlab{b}}){Daddi}, {Elbaz}, {Walter},
  {Bournaud}, {Salmi}, {Carilli}, {Dannerbauer}, {Dickinson}, {Monaco}, \&
  {Riechers}}]{2010ApJ...714L.118D}
{Daddi}, E., {Elbaz}, D., {Walter}, F., {Bournaud}, F., {Salmi}, F., {Carilli},
  C., {Dannerbauer}, H., {Dickinson}, M., {Monaco}, P., \& {Riechers}, D.
  2010{\natexlab{b}}, \apjl, 714, L118

\bibitem[{{Dame} {et~al.}(1986){Dame}, {Elmegreen}, {Cohen}, \&
  {Thaddeus}}]{1986ApJ...305..892D}
{Dame}, T.~M., {Elmegreen}, B.~G., {Cohen}, R.~S., \& {Thaddeus}, P. 1986,
  \apj, 305, 892

\bibitem[{{Draine}(1978)}]{1978ApJS...36..595D}
{Draine}, B.~T. 1978, \apjs, 36, 595

\bibitem[{{Elmegreen}(2000)}]{2000ApJ...530..277E}
{Elmegreen}, B.~G. 2000, \apj, 530, 277

\bibitem[{{Feldmann} \& {Gnedin}(2011)}]{2011ApJ...727L..12F}
{Feldmann}, R., \& {Gnedin}, N.~Y. 2011, \apjl, 727, L12

\bibitem[{{Feldmann} {et~al.}(2011){Feldmann}, {Gnedin}, \&
  {Kravtsov}}]{2011ApJ...732..115F}
{Feldmann}, R., {Gnedin}, N.~Y., \& {Kravtsov}, A.~V. 2011, \apj, 732, 115

\bibitem[{{Feldmann} {et~al.}(2012){Feldmann}, {Gnedin}, \&
  {Kravtsov}}]{2012ApJ...747..124F}
---. 2012, \apj, 747, 124

\bibitem[{{Fu} {et~al.}(2012){Fu}, {Kauffmann}, {Li}, \&
  {Guo}}]{2012arXiv1203.5280F}
{Fu}, J., {Kauffmann}, G., {Li}, C., \& {Guo}, Q. 2012, ArXiv e-prints

\bibitem[{{Gao} \& {Solomon}(2004)}]{2004ApJ...606..271G}
{Gao}, Y., \& {Solomon}, P.~M. 2004, \apj, 606, 271

\bibitem[{{Genzel} {et~al.}(2010){Genzel}, {Tacconi}, {Gracia-Carpio},
  {Sternberg}, {Cooper}, {Shapiro}, {Bolatto}, {Bouch{\'e}}, {Bournaud},
  {Burkert}, {Combes}, {Comerford}, {Cox}, {Davis}, {Schreiber},
  {Garcia-Burillo}, {Lutz}, {Naab}, {Neri}, {Omont}, {Shapley}, \&
  {Weiner}}]{2010MNRAS.407.2091G}
{Genzel}, R., {Tacconi}, L.~J., {Gracia-Carpio}, J., {Sternberg}, A., {Cooper},
  M.~C., {Shapiro}, K., {Bolatto}, A., {Bouch{\'e}}, N., {Bournaud}, F.,
  {Burkert}, A., {Combes}, F., {Comerford}, J., {Cox}, P., {Davis}, M.,
  {Schreiber}, N.~M.~F., {Garcia-Burillo}, S., {Lutz}, D., {Naab}, T., {Neri},
  R., {Omont}, A., {Shapley}, A., \& {Weiner}, B. 2010, \mnras, 407, 2091

\bibitem[{{Glover} \& {Mac Low}(2007)}]{2007ApJ...659.1317G}
{Glover}, S.~C.~O., \& {Mac Low}, M.-M. 2007, \apj, 659, 1317

\bibitem[{{Glover} \& {Mac Low}(2011)}]{2011MNRAS.412..337G}
---. 2011, \mnras, 412, 337

\bibitem[{{Gnedin} \& {Abel}(2001)}]{2001NewA....6..437G}
{Gnedin}, N.~Y., \& {Abel}, T. 2001, New Astronomy, 6, 437

\bibitem[{{Gnedin} \& {Kravtsov}(2010)}]{2010ApJ...714..287G}
{Gnedin}, N.~Y., \& {Kravtsov}, A.~V. 2010, \apj, 714, 287

\bibitem[{{Gnedin} \& {Kravtsov}(2011)}]{2011ApJ...728...88G}
---. 2011, \apj, 728, 88

\bibitem[{{Gnedin} {et~al.}(2009){Gnedin}, {Tassis}, \&
  {Kravtsov}}]{2009ApJ...697...55G}
{Gnedin}, N.~Y., {Tassis}, K., \& {Kravtsov}, A.~V. 2009, \apj, 697, 55

\bibitem[{{Greve} {et~al.}(2005){Greve}, {Bertoldi}, {Smail}, {Neri},
  {Chapman}, {Blain}, {Ivison}, {Genzel}, {Omont}, {Cox}, {Tacconi}, \&
  {Kneib}}]{2005MNRAS.359.1165G}
{Greve}, T.~R., {Bertoldi}, F., {Smail}, I., {Neri}, R., {Chapman}, S.~C.,
  {Blain}, A.~W., {Ivison}, R.~J., {Genzel}, R., {Omont}, A., {Cox}, P.,
  {Tacconi}, L., \& {Kneib}, J.-P. 2005, \mnras, 359, 1165

\bibitem[{{Guibert} {et~al.}(1978){Guibert}, {Lequeux}, \&
  {Viallefond}}]{1978A&A....68....1G}
{Guibert}, J., {Lequeux}, J., \& {Viallefond}, F. 1978, \aap, 68, 1

\bibitem[{{Helfer} {et~al.}(2003){Helfer}, {Thornley}, {Regan}, {Wong},
  {Sheth}, {Vogel}, {Blitz}, \& {Bock}}]{2003ApJS..145..259H}
{Helfer}, T.~T., {Thornley}, M.~D., {Regan}, M.~W., {Wong}, T., {Sheth}, K.,
  {Vogel}, S.~N., {Blitz}, L., \& {Bock}, D.~C.-J. 2003, \apjs, 145, 259

\bibitem[{{Heyer} {et~al.}(2004){Heyer}, {Corbelli}, {Schneider}, \&
  {Young}}]{2004ApJ...602..723H}
{Heyer}, M.~H., {Corbelli}, E., {Schneider}, S.~E., \& {Young}, J.~S. 2004,
  \apj, 602, 723

\bibitem[{{Ivison} {et~al.}(2011){Ivison}, {Papadopoulos}, {Smail}, {Greve},
  {Thomson}, {Xilouris}, \& {Chapman}}]{2011MNRAS.412.1913I}
{Ivison}, R.~J., {Papadopoulos}, P.~P., {Smail}, I., {Greve}, T.~R., {Thomson},
  A.~P., {Xilouris}, E.~M., \& {Chapman}, S.~C. 2011, \mnras, 412, 1913

\bibitem[{{Katz}(1991)}]{1991ApJ...368..325K}
{Katz}, N. 1991, \apj, 368, 325

\bibitem[{{Kennicutt}(1989)}]{1989ApJ...344..685K}
{Kennicutt}, Jr., R.~C. 1989, \apj, 344, 685

\bibitem[{{Kennicutt}(1998)}]{1998ApJ...498..541K}
---. 1998, \apj, 498, 541

\bibitem[{{Kennicutt} {et~al.}(2007){Kennicutt}, {Calzetti}, {Walter}, {Helou},
  {Hollenbach}, {Armus}, {Bendo}, {Dale}, {Draine}, {Engelbracht}, {Gordon},
  {Prescott}, {Regan}, {Thornley}, {Bot}, {Brinks}, {de Blok}, {de Mello},
  {Meyer}, {Moustakas}, {Murphy}, {Sheth}, \& {Smith}}]{2007ApJ...671..333K}
{Kennicutt}, Jr., R.~C., {Calzetti}, D., {Walter}, F., {Helou}, G.,
  {Hollenbach}, D.~J., {Armus}, L., {Bendo}, G., {Dale}, D.~A., {Draine},
  B.~T., {Engelbracht}, C.~W., {Gordon}, K.~D., {Prescott}, M.~K.~M., {Regan},
  M.~W., {Thornley}, M.~D., {Bot}, C., {Brinks}, E., {de Blok}, E., {de Mello},
  D., {Meyer}, M., {Moustakas}, J., {Murphy}, E.~J., {Sheth}, K., \& {Smith},
  J.~D.~T. 2007, \apj, 671, 333

\bibitem[{{Komugi} {et~al.}(2005){Komugi}, {Sofue}, {Nakanishi}, {Onodera}, \&
  {Egusa}}]{2005PASJ...57..733K}
{Komugi}, S., {Sofue}, Y., {Nakanishi}, H., {Onodera}, S., \& {Egusa}, F. 2005,
  \pasj, 57, 733

\bibitem[{{Kravtsov} {et~al.}(2002){Kravtsov}, {Klypin}, \&
  {Hoffman}}]{2002ApJ...571..563K}
{Kravtsov}, A.~V., {Klypin}, A., \& {Hoffman}, Y. 2002, \apj, 571, 563

\bibitem[{{Kravtsov} {et~al.}(1997){Kravtsov}, {Klypin}, \&
  {Khokhlov}}]{1997ApJS..111...73K}
{Kravtsov}, A.~V., {Klypin}, A.~A., \& {Khokhlov}, A.~M. 1997, \apjs, 111, 73

\bibitem[{{Krumholz} {et~al.}(2011){Krumholz}, {Leroy}, \&
  {McKee}}]{2011ApJ...731...25K}
{Krumholz}, M.~R., {Leroy}, A.~K., \& {McKee}, C.~F. 2011, \apj, 731, 25

\bibitem[{{Krumholz} \& {McKee}(2005)}]{2005ApJ...630..250K}
{Krumholz}, M.~R., \& {McKee}, C.~F. 2005, \apj, 630, 250

\bibitem[{{Krumholz} {et~al.}(2009){Krumholz}, {McKee}, \&
  {Tumlinson}}]{2009ApJ...699..850K}
{Krumholz}, M.~R., {McKee}, C.~F., \& {Tumlinson}, J. 2009, \apj, 699, 850

\bibitem[{{Krumholz} \& {Thompson}(2007)}]{2007ApJ...669..289K}
{Krumholz}, M.~R., \& {Thompson}, T.~A. 2007, \apj, 669, 289

\bibitem[{{Kuhlen} {et~al.}(2012){Kuhlen}, {Krumholz}, {Madau}, {Smith}, \&
  {Wise}}]{2012ApJ...749...36K}
{Kuhlen}, M., {Krumholz}, M.~R., {Madau}, P., {Smith}, B.~D., \& {Wise}, J.
  2012, \apj, 749, 36

\bibitem[{{Kuno} {et~al.}(2007){Kuno}, {Sato}, {Nakanishi}, {Hirota}, {Tosaki},
  {Shioya}, {Sorai}, {Nakai}, {Nishiyama}, \&
  {Vila-Vilar{\'o}}}]{2007PASJ...59..117K}
{Kuno}, N., {Sato}, N., {Nakanishi}, H., {Hirota}, A., {Tosaki}, T., {Shioya},
  Y., {Sorai}, K., {Nakai}, N., {Nishiyama}, K., \& {Vila-Vilar{\'o}}, B. 2007,
  \pasj, 59, 117

\bibitem[{{Lada} {et~al.}(2012){Lada}, {Forbrich}, {Lombardi}, \&
  {Alves}}]{2012ApJ...745..190L}
{Lada}, C.~J., {Forbrich}, J., {Lombardi}, M., \& {Alves}, J.~F. 2012, \apj,
  745, 190

\bibitem[{{Lada} \& {Lada}(2003)}]{2003ARA&A..41...57L}
{Lada}, C.~J., \& {Lada}, E.~A. 2003, \araa, 41, 57

\bibitem[{{Lada} {et~al.}(2010){Lada}, {Lombardi}, \&
  {Alves}}]{2010ApJ...724..687L}
{Lada}, C.~J., {Lombardi}, M., \& {Alves}, J.~F. 2010, \apj, 724, 687

\bibitem[{{Lagos} {et~al.}(2012){Lagos}, {Bayet}, {Baugh}, {Lacey}, {Bell},
  {Fanidakis}, \& {Geach}}]{2012arXiv1204.0795L}
{Lagos}, C.~d.~P., {Bayet}, E., {Baugh}, C.~M., {Lacey}, C.~G., {Bell}, T.,
  {Fanidakis}, N., \& {Geach}, J. 2012, ArXiv e-prints

\bibitem[{{Larsen}(2002)}]{2002AJ....124.1393L}
{Larsen}, S.~S. 2002, \aj, 124, 1393

\bibitem[{{Larson}(1981)}]{1981MNRAS.194..809L}
{Larson}, R.~B. 1981, \mnras, 194, 809

\bibitem[{{Leitherer} {et~al.}(2010){Leitherer}, {Ortiz Ot{\'a}lvaro},
  {Bresolin}, {Kudritzki}, {Lo Faro}, {Pauldrach}, {Pettini}, \&
  {Rix}}]{2010ApJS..189..309L}
{Leitherer}, C., {Ortiz Ot{\'a}lvaro}, P.~A., {Bresolin}, F., {Kudritzki},
  R.-P., {Lo Faro}, B., {Pauldrach}, A.~W.~A., {Pettini}, M., \& {Rix}, S.~A.
  2010, \apjs, 189, 309

\bibitem[{{Leroy} {et~al.}(2012){Leroy}, {Bigiel}, {de Blok}, {Boissier},
  {Bolatto}, {Brinks}, {Madore}, {Munoz-Mateos}, {Murphy}, {Sandstrom},
  {Schruba}, \& {Walter}}]{2012arXiv1202.2873L}
{Leroy}, A.~K., {Bigiel}, F., {de Blok}, W.~J.~G., {Boissier}, S., {Bolatto},
  A., {Brinks}, E., {Madore}, B., {Munoz-Mateos}, J.-C., {Murphy}, E.,
  {Sandstrom}, K., {Schruba}, A., \& {Walter}, F. 2012, ArXiv e-prints

\bibitem[{{Liu} {et~al.}(2011){Liu}, {Koda}, {Calzetti}, {Fukuhara}, \&
  {Momose}}]{2011ApJ...735...63L}
{Liu}, G., {Koda}, J., {Calzetti}, D., {Fukuhara}, M., \& {Momose}, R. 2011,
  \apj, 735, 63

\bibitem[{{Maiolino} {et~al.}(2008){Maiolino}, {Nagao}, {Grazian}, {Cocchia},
  {Marconi}, {Mannucci}, {Cimatti}, {Pipino}, {Ballero}, {Calura}, {Chiappini},
  {Fontana}, {Granato}, {Matteucci}, {Pastorini}, {Pentericci}, {Risaliti},
  {Salvati}, \& {Silva}}]{2008A&A...488..463M}
{Maiolino}, R., {Nagao}, T., {Grazian}, A., {Cocchia}, F., {Marconi}, A.,
  {Mannucci}, F., {Cimatti}, A., {Pipino}, A., {Ballero}, S., {Calura}, F.,
  {Chiappini}, C., {Fontana}, A., {Granato}, G.~L., {Matteucci}, F.,
  {Pastorini}, G., {Pentericci}, L., {Risaliti}, G., {Salvati}, M., \& {Silva},
  L. 2008, \aap, 488, 463

\bibitem[{{Mathis} {et~al.}(1983){Mathis}, {Mezger}, \&
  {Panagia}}]{1983A&A...128..212M}
{Mathis}, J.~S., {Mezger}, P.~G., \& {Panagia}, N. 1983, \aap, 128, 212

\bibitem[{{McKee} \& {Ostriker}(2007)}]{2007ARA&A..45..565M}
{McKee}, C.~F., \& {Ostriker}, E.~C. 2007, \araa, 45, 565

\bibitem[{{Momose} {et~al.}(2010){Momose}, {Okumura}, {Koda}, \&
  {Sawada}}]{2010ApJ...721..383M}
{Momose}, R., {Okumura}, S.~K., {Koda}, J., \& {Sawada}, T. 2010, \apj, 721,
  383

\bibitem[{{Morrissey} {et~al.}(2005){Morrissey}, {Schiminovich}, {Barlow},
  {Martin}, {Blakkolb}, {Conrow}, {Cooke}, {Erickson}, {Fanson}, {Friedman},
  {Grange}, {Jelinsky}, {Lee}, {Liu}, {Mazer}, {McLean}, {Milliard}, {Randall},
  {Schmitigal}, {Sen}, {Siegmund}, {Surber}, {Vaughan}, {Viton}, {Welsh},
  {Bianchi}, {Byun}, {Donas}, {Forster}, {Heckman}, {Lee}, {Madore}, {Malina},
  {Neff}, {Rich}, {Small}, {Szalay}, \& {Wyder}}]{2005ApJ...619L...7M}
{Morrissey}, P., {Schiminovich}, D., {Barlow}, T.~A., {Martin}, D.~C.,
  {Blakkolb}, B., {Conrow}, T., {Cooke}, B., {Erickson}, K., {Fanson}, J.,
  {Friedman}, P.~G., {Grange}, R., {Jelinsky}, P.~N., {Lee}, S.-C., {Liu}, D.,
  {Mazer}, A., {McLean}, R., {Milliard}, B., {Randall}, D., {Schmitigal}, W.,
  {Sen}, A., {Siegmund}, O.~H.~W., {Surber}, F., {Vaughan}, A., {Viton}, M.,
  {Welsh}, B.~Y., {Bianchi}, L., {Byun}, Y.-I., {Donas}, J., {Forster}, K.,
  {Heckman}, T.~M., {Lee}, Y.-W., {Madore}, B.~F., {Malina}, R.~F., {Neff},
  S.~G., {Rich}, R.~M., {Small}, T., {Szalay}, A.~S., \& {Wyder}, T.~K. 2005,
  \apjl, 619, L7

\bibitem[{{Murray}(2011)}]{2011ApJ...729..133M}
{Murray}, N. 2011, \apj, 729, 133

\bibitem[{{Narayanan} {et~al.}(2011{\natexlab{a}}){Narayanan}, {Cox},
  {Hayward}, \& {Hernquist}}]{2011MNRAS.412..287N}
{Narayanan}, D., {Cox}, T.~J., {Hayward}, C.~C., \& {Hernquist}, L.
  2011{\natexlab{a}}, \mnras, 412, 287

\bibitem[{{Narayanan} {et~al.}(2011{\natexlab{b}}){Narayanan}, {Krumholz},
  {Ostriker}, \& {Hernquist}}]{2011MNRAS.418..664N}
{Narayanan}, D., {Krumholz}, M., {Ostriker}, E.~C., \& {Hernquist}, L.
  2011{\natexlab{b}}, \mnras, 418, 664

\bibitem[{{Narayanan} {et~al.}(2012){Narayanan}, {Krumholz}, {Ostriker}, \&
  {Hernquist}}]{2012MNRAS.tmp.2537N}
{Narayanan}, D., {Krumholz}, M.~R., {Ostriker}, E.~C., \& {Hernquist}, L. 2012,
  \mnras, 2537

\bibitem[{{Onodera} {et~al.}(2010){Onodera}, {Kuno}, {Tosaki}, {Kohno},
  {Nakanishi}, {Sawada}, {Muraoka}, {Komugi}, {Miura}, {Kaneko}, {Hirota}, \&
  {Kawabe}}]{2010ApJ...722L.127O}
{Onodera}, S., {Kuno}, N., {Tosaki}, T., {Kohno}, K., {Nakanishi}, K.,
  {Sawada}, T., {Muraoka}, K., {Komugi}, S., {Miura}, R., {Kaneko}, H.,
  {Hirota}, A., \& {Kawabe}, R. 2010, \apjl, 722, L127

\bibitem[{{Papadopoulos} \& {Pelupessy}(2010)}]{2010ApJ...717.1037P}
{Papadopoulos}, P.~P., \& {Pelupessy}, F.~I. 2010, \apj, 717, 1037

\bibitem[{{Papadopoulos} {et~al.}(2012){Papadopoulos}, {van der Werf},
  {Xilouris}, {Isaak}, \& {Gao}}]{2012arXiv1202.1803P}
{Papadopoulos}, P.~P., {van der Werf}, P., {Xilouris}, E., {Isaak}, K.~G., \&
  {Gao}, Y. 2012, ArXiv e-prints

\bibitem[{{Pelupessy} {et~al.}(2006){Pelupessy}, {Papadopoulos}, \& {van der
  Werf}}]{2006ApJ...645.1024P}
{Pelupessy}, F.~I., {Papadopoulos}, P.~P., \& {van der Werf}, P. 2006, \apj,
  645, 1024

\bibitem[{{Portegies Zwart} {et~al.}(2010){Portegies Zwart}, {McMillan}, \&
  {Gieles}}]{2010ARA&A..48..431P}
{Portegies Zwart}, S.~F., {McMillan}, S.~L.~W., \& {Gieles}, M. 2010, \araa,
  48, 431

\bibitem[{{Rahman} {et~al.}(2011){Rahman}, {Bolatto}, {Wong}, {Leroy},
  {Walter}, {Rosolowsky}, {West}, {Bigiel}, {Ott}, {Xue}, {Herrera-Camus},
  {Jameson}, {Blitz}, \& {Vogel}}]{2011ApJ...730...72R}
{Rahman}, N., {Bolatto}, A.~D., {Wong}, T., {Leroy}, A.~K., {Walter}, F.,
  {Rosolowsky}, E., {West}, A.~A., {Bigiel}, F., {Ott}, J., {Xue}, R.,
  {Herrera-Camus}, R., {Jameson}, K., {Blitz}, L., \& {Vogel}, S.~N. 2011,
  \apj, 730, 72

\bibitem[{{Rahman} {et~al.}(2012){Rahman}, {Bolatto}, {Xue}, {Wong}, {Leroy},
  {Walter}, {Bigiel}, {Rosolowsky}, {Fisher}, {Vogel}, {Blitz}, {West}, \&
  {Ott}}]{2012ApJ...745..183R}
{Rahman}, N., {Bolatto}, A.~D., {Xue}, R., {Wong}, T., {Leroy}, A.~K.,
  {Walter}, F., {Bigiel}, F., {Rosolowsky}, E., {Fisher}, D.~B., {Vogel},
  S.~N., {Blitz}, L., {West}, A.~A., \& {Ott}, J. 2012, \apj, 745, 183

\bibitem[{{Regan} {et~al.}(2001){Regan}, {Thornley}, {Helfer}, {Sheth}, {Wong},
  {Vogel}, {Blitz}, \& {Bock}}]{2001ApJ...561..218R}
{Regan}, M.~W., {Thornley}, M.~D., {Helfer}, T.~T., {Sheth}, K., {Wong}, T.,
  {Vogel}, S.~N., {Blitz}, L., \& {Bock}, D.~C.-J. 2001, \apj, 561, 218

\bibitem[{{Riechers} {et~al.}(2006){Riechers}, {Walter}, {Carilli}, {Knudsen},
  {Lo}, {Benford}, {Staguhn}, {Hunter}, {Bertoldi}, {Henkel}, {Menten},
  {Weiss}, {Yun}, \& {Scoville}}]{2006ApJ...650..604R}
{Riechers}, D.~A., {Walter}, F., {Carilli}, C.~L., {Knudsen}, K.~K., {Lo},
  K.~Y., {Benford}, D.~J., {Staguhn}, J.~G., {Hunter}, T.~R., {Bertoldi}, F.,
  {Henkel}, C., {Menten}, K.~M., {Weiss}, A., {Yun}, M.~S., \& {Scoville},
  N.~Z. 2006, \apj, 650, 604

\bibitem[{{Robertson} \& {Kravtsov}(2008)}]{2008ApJ...680.1083R}
{Robertson}, B.~E., \& {Kravtsov}, A.~V. 2008, \apj, 680, 1083

\bibitem[{{Saintonge} {et~al.}(2011){Saintonge}, {Kauffmann}, {Wang}, {Kramer},
  {Tacconi}, {Buchbender}, {Catinella}, {Graci{\'a}-Carpio}, {Cortese},
  {Fabello}, {Fu}, {Genzel}, {Giovanelli}, {Guo}, {Haynes}, {Heckman},
  {Krumholz}, {Lemonias}, {Li}, {Moran}, {Rodriguez-Fernandez}, {Schiminovich},
  {Schuster}, \& {Sievers}}]{2011MNRAS.415...61S}
{Saintonge}, A., {Kauffmann}, G., {Wang}, J., {Kramer}, C., {Tacconi}, L.~J.,
  {Buchbender}, C., {Catinella}, B., {Graci{\'a}-Carpio}, J., {Cortese}, L.,
  {Fabello}, S., {Fu}, J., {Genzel}, R., {Giovanelli}, R., {Guo}, Q., {Haynes},
  M.~P., {Heckman}, T.~M., {Krumholz}, M.~R., {Lemonias}, J., {Li}, C.,
  {Moran}, S., {Rodriguez-Fernandez}, N., {Schiminovich}, D., {Schuster}, K.,
  \& {Sievers}, A. 2011, \mnras, 415, 61

\bibitem[{{Schaye} \& {Dalla Vecchia}(2008)}]{2008MNRAS.383.1210S}
{Schaye}, J., \& {Dalla Vecchia}, C. 2008, \mnras, 383, 1210

\bibitem[{{Schmidt}(1959)}]{1959ApJ...129..243S}
{Schmidt}, M. 1959, \apj, 129, 243

\bibitem[{{Schruba} {et~al.}(2011){Schruba}, {Leroy}, {Walter}, {Bigiel},
  {Brinks}, {de Blok}, {Dumas}, {Kramer}, {Rosolowsky}, {Sandstrom},
  {Schuster}, {Usero}, {Weiss}, \& {Wiesemeyer}}]{2011AJ....142...37S}
{Schruba}, A., {Leroy}, A.~K., {Walter}, F., {Bigiel}, F., {Brinks}, E., {de
  Blok}, W.~J.~G., {Dumas}, G., {Kramer}, C., {Rosolowsky}, E., {Sandstrom},
  K., {Schuster}, K., {Usero}, A., {Weiss}, A., \& {Wiesemeyer}, H. 2011, \aj,
  142, 37

\bibitem[{{Schruba} {et~al.}(2010){Schruba}, {Leroy}, {Walter}, {Sandstrom}, \&
  {Rosolowsky}}]{2010ApJ...722.1699S}
{Schruba}, A., {Leroy}, A.~K., {Walter}, F., {Sandstrom}, K., \& {Rosolowsky},
  E. 2010, \apj, 722, 1699

\bibitem[{{Scoville} \& {Sanders}(1987)}]{1987ASSL..134...21S}
{Scoville}, N.~Z., \& {Sanders}, D.~B. 1987, in Astrophysics and Space Science
  Library, Vol. 134, Interstellar Processes, ed. {D.~J.~Hollenbach \&
  H.~A.~Thronson Jr.}, 21--50

\bibitem[{{Searle}(1971)}]{1971ApJ...168..327S}
{Searle}, L. 1971, \apj, 168, 327

\bibitem[{{Shetty} {et~al.}(2011){Shetty}, {Glover}, {Dullemond}, \&
  {Klessen}}]{2011MNRAS.412.1686S}
{Shetty}, R., {Glover}, S.~C., {Dullemond}, C.~P., \& {Klessen}, R.~S. 2011,
  \mnras, 412, 1686

\bibitem[{{Solomon} {et~al.}(1987){Solomon}, {Rivolo}, {Barrett}, \&
  {Yahil}}]{1987ApJ...319..730S}
{Solomon}, P.~M., {Rivolo}, A.~R., {Barrett}, J., \& {Yahil}, A. 1987, \apj,
  319, 730

\bibitem[{{Solomon} \& {Vanden Bout}(2005)}]{2005ARA&A..43..677S}
{Solomon}, P.~M., \& {Vanden Bout}, P.~A. 2005, \araa, 43, 677

\bibitem[{{Tacconi} {et~al.}(2010){Tacconi}, {Genzel}, {Neri}, {Cox}, {Cooper},
  {Shapiro}, {Bolatto}, {Bouch{\'e}}, {Bournaud}, {Burkert}, {Combes},
  {Comerford}, {Davis}, {Schreiber}, {Garcia-Burillo}, {Gracia-Carpio}, {Lutz},
  {Naab}, {Omont}, {Shapley}, {Sternberg}, \& {Weiner}}]{2010Natur.463..781T}
{Tacconi}, L.~J., {Genzel}, R., {Neri}, R., {Cox}, P., {Cooper}, M.~C.,
  {Shapiro}, K., {Bolatto}, A., {Bouch{\'e}}, N., {Bournaud}, F., {Burkert},
  A., {Combes}, F., {Comerford}, J., {Davis}, M., {Schreiber}, N.~M.~F.,
  {Garcia-Burillo}, S., {Gracia-Carpio}, J., {Lutz}, D., {Naab}, T., {Omont},
  A., {Shapley}, A., {Sternberg}, A., \& {Weiner}, B. 2010, \nat, 463, 781

\bibitem[{{Tacconi} {et~al.}(2006){Tacconi}, {Neri}, {Chapman}, {Genzel},
  {Smail}, {Ivison}, {Bertoldi}, {Blain}, {Cox}, {Greve}, \&
  {Omont}}]{2006ApJ...640..228T}
{Tacconi}, L.~J., {Neri}, R., {Chapman}, S.~C., {Genzel}, R., {Smail}, I.,
  {Ivison}, R.~J., {Bertoldi}, F., {Blain}, A., {Cox}, P., {Greve}, T., \&
  {Omont}, A. 2006, \apj, 640, 228

\bibitem[{{Talbot}(1971)}]{1971ApL.....8..111T}
{Talbot}, Jr., R.~J. 1971, \aplett, 8, 111

\bibitem[{{Talbot}(1980)}]{1980ApJ...235..821T}
---. 1980, \apj, 235, 821

\bibitem[{{Tasker}(2011)}]{2011ApJ...730...11T}
{Tasker}, E.~J. 2011, \apj, 730, 11

\bibitem[{{Tasker} \& {Tan}(2009)}]{2009ApJ...700..358T}
{Tasker}, E.~J., \& {Tan}, J.~C. 2009, \apj, 700, 358

\bibitem[{{Teyssier} {et~al.}(2010){Teyssier}, {Chapon}, \&
  {Bournaud}}]{2010ApJ...720L.149T}
{Teyssier}, R., {Chapon}, D., \& {Bournaud}, F. 2010, \apjl, 720, L149

\bibitem[{{Thilker} {et~al.}(2007){Thilker}, {Boissier}, {Bianchi}, {Calzetti},
  {Boselli}, {Dale}, {Seibert}, {Braun}, {Burgarella}, {Gil de Paz}, {Helou},
  {Walter}, {Kennicutt}, {Madore}, {Martin}, {Barlow}, {Forster}, {Friedman},
  {Morrissey}, {Neff}, {Schiminovich}, {Small}, {Wyder}, {Donas}, {Heckman},
  {Lee}, {Milliard}, {Rich}, {Szalay}, {Welsh}, \& {Yi}}]{2007ApJS..173..572T}
{Thilker}, D.~A., {Boissier}, S., {Bianchi}, L., {Calzetti}, D., {Boselli}, A.,
  {Dale}, D.~A., {Seibert}, M., {Braun}, R., {Burgarella}, D., {Gil de Paz},
  A., {Helou}, G., {Walter}, F., {Kennicutt}, Jr., R.~C., {Madore}, B.~F.,
  {Martin}, D.~C., {Barlow}, T.~A., {Forster}, K., {Friedman}, P.~G.,
  {Morrissey}, P., {Neff}, S.~G., {Schiminovich}, D., {Small}, T., {Wyder},
  T.~K., {Donas}, J., {Heckman}, T.~M., {Lee}, Y.-W., {Milliard}, B., {Rich},
  R.~M., {Szalay}, A.~S., {Welsh}, B.~Y., \& {Yi}, S.~K. 2007, \apjs, 173, 572

\bibitem[{{Verley} {et~al.}(2010){Verley}, {Corbelli}, {Giovanardi}, \&
  {Hunt}}]{2010A&A...510A..64V}
{Verley}, S., {Corbelli}, E., {Giovanardi}, C., \& {Hunt}, L.~K. 2010, \aap,
  510, A64

\bibitem[{{Weidner} {et~al.}(2004){Weidner}, {Kroupa}, \&
  {Larsen}}]{2004MNRAS.350.1503W}
{Weidner}, C., {Kroupa}, P., \& {Larsen}, S.~S. 2004, \mnras, 350, 1503

\bibitem[{{Wilson} {et~al.}(1970){Wilson}, {Jefferts}, \&
  {Penzias}}]{1970ApJ...161L..43W}
{Wilson}, R.~W., {Jefferts}, K.~B., \& {Penzias}, A.~A. 1970, \apjl, 161, L43+

\bibitem[{{Wolfire} {et~al.}(2010){Wolfire}, {Hollenbach}, \&
  {McKee}}]{2010ApJ...716.1191W}
{Wolfire}, M.~G., {Hollenbach}, D., \& {McKee}, C.~F. 2010, \apj, 716, 1191

\bibitem[{{Wong} \& {Blitz}(2002)}]{2002ApJ...569..157W}
{Wong}, T., \& {Blitz}, L. 2002, \apj, 569, 157

\bibitem[{{Young} {et~al.}(1995){Young}, {Xie}, {Tacconi}, {Knezek}, {Viscuso},
  {Tacconi-Garman}, {Scoville}, {Schneider}, {Schloerb}, {Lord}, {Lesser},
  {Kenney}, {Huang}, {Devereux}, {Claussen}, {Case}, {Carpenter}, {Berry}, \&
  {Allen}}]{1995ApJS...98..219Y}
{Young}, J.~S., {Xie}, S., {Tacconi}, L., {Knezek}, P., {Viscuso}, P.,
  {Tacconi-Garman}, L., {Scoville}, N., {Schneider}, S., {Schloerb}, F.~P.,
  {Lord}, S., {Lesser}, A., {Kenney}, J., {Huang}, Y.-L., {Devereux}, N.,
  {Claussen}, M., {Case}, J., {Carpenter}, J., {Berry}, M., \& {Allen}, L.
  1995, \apjs, 98, 219

\bibitem[{{Zemp} {et~al.}(2012){Zemp}, {Gnedin}, {Gnedin}, \&
  {Kravtsov}}]{2012ApJ...748...54Z}
{Zemp}, M., {Gnedin}, O.~Y., {Gnedin}, N.~Y., \& {Kravtsov}, A.~V. 2012, \apj,
  748, 54

\bibitem[{{Zuckerman} \& {Evans}(1974)}]{1974ApJ...192L.149Z}
{Zuckerman}, B., \& {Evans}, II, N.~J. 1974, \apjl, 192, L149

\end{thebibliography}

\end{document}